\documentclass[a4paper, amsfonts, amssymb, amsmath, reprint, showkeys, showpacs, nofootinbib, twoside, superscriptaddress,aps,prd]{revtex4-1}
\usepackage[english]{babel}
\usepackage[utf8]{inputenc}
\usepackage{comment}
\usepackage[colorinlistoftodos, color=green!40, prependcaption]{todonotes}
\usepackage{appendix}
\usepackage{amsthm}
\usepackage{mathtools}
\usepackage{physics}
\usepackage{xcolor}
\usepackage{graphicx}
\usepackage{adjustbox}
\usepackage{placeins}
\usepackage[T1]{fontenc}
\usepackage{lipsum}
\usepackage{csquotes}
\usepackage{orcidlink}
\usepackage[caption=false]{subfig}

\usepackage{ulem}
\usepackage{hyperref}

\begin{document}
\title{Gravitational lensing by extremal rotating black holes in the strong deflection limit}

    \author{Fabiano Feleppa\,\orcidlink{0000-0002-3706-1021}}
    \email[]{ffeleppa@unisa.it}
     \affiliation{Dipartimento di Fisica “E.R. Caianiello”, Università di Salerno, Via Giovanni Paolo II 132, I-84084 Fisciano, Italy}
    \affiliation{Istituto Nazionale di Fisica Nucleare, Sezione di Napoli, Via Cintia, 80126, Napoli, Italy}
    \affiliation{Center of Gravity, Niels Bohr Institute, Blegdamsvej 17, 2100 Copenhagen, Denmark}

    \author{Valerio Bozza\,\orcidlink{0000-0003-4590-0136}}
    \email[]{vbozza@unisa.it}
    \affiliation{Dipartimento di Fisica “E.R. Caianiello”, Università di Salerno, Via Giovanni Paolo II 132, I-84084 Fisciano, Italy}
    \affiliation{Istituto Nazionale di Fisica Nucleare, Sezione di Napoli, Via Cintia, 80126, Napoli, Italy}

    \author{Welmoed Marit de Graaf\,\orcidlink{0009-0002-1662-7441}}
    \email[]{w.degraaf@studenti.unisa.it}
    \affiliation{Dipartimento di Fisica “E.R. Caianiello”, Università di Salerno, Via Giovanni Paolo II 132, I-84084 Fisciano, Italy}

\begin{abstract}
In the strong deflection regime, light rays passing close to an astrophysical black hole may remain trapped near unstable photon orbits for a long time before escaping to infinity. The traditional strong-deflection limit, which accurately describes the logarithmic divergence of the deflection angle for spherically symmetric and slowly rotating black holes, breaks down when the relevant prograde critical photon orbit coincides with the degenerate horizon of an extremal rotating black hole. We present a new strong deflection limit expansion for this horizon critical orbit, covering a general class of extremal rotating black holes. We show that the deflection angle exhibits a stronger power-law divergence in addition to the logarithmic divergence. For an adequate description of higher-order images, additional terms in the expansion must be retained. We first study prograde gravitational lensing in the equatorial plane and then extend the analysis to quasi-equatorial motion, which allows us to calculate the magnification of the higher-order images and the position of the caustic points. We finally apply the general framework to explicit examples, including the Kerr, Kerr-Newman, and Kerr-Sen metrics.
\end{abstract}


\maketitle

\section{Introduction}

The study of light bending by compact objects has gained considerable interest with the advent of horizon-scale observations of astrophysical black holes by the Event Horizon Telescope \cite{L1,L2,L3,L4,L5,L6, Kocherlakota-2021,L12,L13,L14,L15,L16,L17}, which have opened the possibility of probing lensing beyond the weak-deflection approximation. From a theoretical perspective, however, the strong bending of light by black holes has a much longer history, rooted in the analysis of photon trajectories close to unstable circular orbits.

The first exact treatment of light deflection in the Schwarzschild spacetime was given by Darwin \cite{Darwin1959}. The expansion of Darwin's exact result near the critical photon orbit leads to a logarithmic approximation that is now known as the strong deflection limit. This approximation describes photon trajectories whose total azimuthal shift can become arbitrarily large as the distance of closest approach tends to the photon-sphere radius. In this regime, photons can wind around the black hole one or more times before escaping to infinity, producing multiple images of the same source; these images appear on either side of the black hole and are often referred to as higher-order images or relativistic images. After Darwin's pioneering work, several aspects of black-hole lensing and higher-order images were discussed by Atkinson \cite{Atkinson1965}, Misner, Thorne and Wheeler \cite{Misner1973}, Luminet \cite{Luminet1979}, and Ohanian \cite{Ohanian1987}. The properties of higher-order images were later computed numerically for the Schwarzschild black hole by Virbhadra and Ellis \cite{Ellis2000}, while exact lens-equation approaches were developed by Frittelli, Kling and Newman \cite{Frittelli2000} and by Perlick \cite{Perlick2004}. The strong deflection limit was then formulated as a systematic analytical tool for calculating, e.g., the positions and magnifications of higher-order images, first for Schwarzschild black holes \cite{Bozza2001}, then for generic static and spherically symmetric spacetimes \cite{Bozza2002}, and subsequently for rotating black holes \cite{Bozza2003,Bozza2005}. Its range of applicability has also been enlarged to include photon propagation in inhomogeneous plasma \cite{Feleppaplasma2024} and massive-particle deflection in spherically symmetric spacetimes \cite{Feleppamassiveparticles2025}. These developments have motivated a large body of work on lensing in the strong-deflection regime \cite{Claudel2001,Hasse2002,Perlick2004-review,Iyer2007,Keeton2008,Tsupko2008, Majumdar2009,Tarasenko2010,Eiroa2011, Wei2012,Zhang2015,Alhamzawi2016,Tsukamoto2016,Aldi-Bozza-2017,Dai2018, Aratore2021,Kuang2022,Abbas2023,Duan2023,Aratore-Bozza-2024,Vachher2025,Feleppa2025, Tsukamoto2026,Igata20261,Igata20262,Igata20263}, including studies of higher-order images in the form of photon rings \cite{Gralla2019,Johnson-2020,Gralla2020,Lupsasca2020,Gralla-Lupsasca-2020,Wielgus-2021,Broderick-2022,Ayzenberg-2022,Guerrero-2022,BK-Tsupko-2022,Tsupko-2022,Eichhorn-2023, Broderick-Salehi-2023,Kocherlakota-2024-1,Kocherlakota-2024-2,Aratore-Tsupko-Perlick-2024,Tsupko2025}.

The strong deflection limit construction, however, has an important limitation in the context of rapidly rotating black holes. In Kerr lensing, for prograde photon motion, the coefficients that control the leading near-critical divergence and the finite part of the deflection angle become singular as the spin approaches extremality, as shown in Fig.~3 of Ref.~\cite{Bozza2003}. This signals that a different asymptotic treatment is required. The origin of this breakdown is geometric:~in the extremal Kerr limit, the prograde unstable circular photon orbit coincides with the degenerate event horizon. As a consequence, photons in the strong-deflection regime can approach the horizon arbitrarily closely before escaping, and the singularity structure of the deflection integral changes. One is therefore led to expect a stronger divergence than the logarithmic one characterizing the non-extremal case. High-spin black-hole lensing has also been studied from complementary perspectives in Refs.~\cite{Gralla2018,Chen2025}.

There is also observational motivation for developing analytical tools that remain well defined in the extremal or near-extremal regime. Indeed, an analysis based on the twisted-light technique, under the assumption that M$87^*$ is described by the Kerr metric, suggests that it is a rapidly rotating black hole \cite{Tamburini1,Tamburini2,Tamburini3}.

\begin{figure}[t!]
    \centering \includegraphics[width=1.0\linewidth]{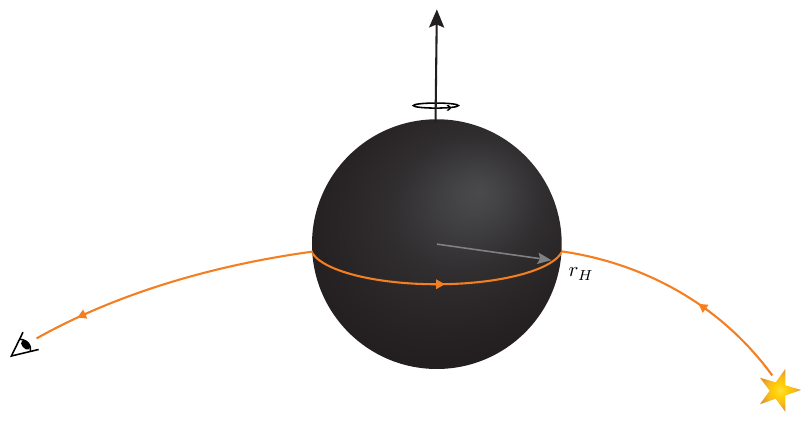}
    \caption{Schematic representation of the extremal prograde lensing configuration considered in this work. The black surface denotes the degenerate event horizon, whose radius $r_H$ coincides with the critical radius of the prograde circular photon orbit. The orange curve represents a photon trajectory in the equatorial plane that winds close to the horizon before escaping to a distant observer.} \label{fig:lensingconfiguration}
\end{figure}

In this work we address this problem for exact extremal rotating black holes, starting from photon motion confined to the equatorial plane. We develop a strong-deflection expansion adapted to the case in which the prograde critical photon orbit lies on the horizon. The lensing configuration of interest is shown schematically in Fig.~\ref{fig:lensingconfiguration}.

The paper is organized as follows. In Sec.~\ref{sec:prelanddeflangle}, we introduce the class of extremal Kerr-like spacetimes considered in this work and recall the exact expression for the deflection angle of equatorial photon trajectories. In Sec.~\ref{sec:sdlanalysis}, we derive the extremal strong-deflection expansion. We first isolate the leading power-law divergence, then compute the subleading logarithmic term, add the regular contribution, and finally specialize the result to the extremal Kerr metric. In Sec.~\ref{sec:higherorderimages}, we use this expansion to calculate the angular positions of higher-order images. In Sec.~\ref{sec:linearcontributions}, we improve the approximation by retaining terms linear in the distance from the horizon and apply the improved expansion to the extremal Kerr case. In Sec.~\ref{sec:KNKS}, we apply the formalism to other extremal rotating geometries, namely the Kerr-Newman and Kerr-Sen black holes. In Sec.~\ref{sec:smalldeclinations}, we extend the analysis to quasi-equatorial photon motion and derive the corresponding polar phase in the strong-deflection limit. In Sec.~\ref{sec:magncaustics}, we use the quasi-equatorial lens map to compute magnifications and determine the caustic points. Finally, Sec.~\ref{sec:conclusions} is devoted to conclusions and possible future developments.

\section{Extremal Kerr-like spacetimes and deflection angle of photons}
\label{sec:prelanddeflangle}

Here we restrict attention to a Kerr-like subclass of stationary,
axisymmetric, and asymptotically flat black-hole spacetimes admitting
Boyer-Lindquist-type coordinates $x^\mu=(t,r,\vartheta,\varphi)$. In
what follows, we set $G=c=1$ and work with signature convention
$\{+,-,-,-\}$. Considering light rays moving on the equatorial plane
$(\vartheta=\pi/2)$, the metric takes the form \cite{Bozza2003}
\begin{multline}
    \mathrm{d}s^2 = g_{\mu\nu}\,\mathrm{d}x^\mu\,\mathrm{d}x^\nu
    = A(r)\,\mathrm{d}t^2 - B(r)\,\mathrm{d}r^2
    \\
    - C(r)\,\mathrm{d}\varphi^2 + 2D(r)\,\mathrm{d}t\,\mathrm{d}\varphi\,.
\label{genericmetric}
\end{multline}
Since null geodesics are invariant under conformal rescalings of the metric, we can exploit this freedom. For the rotating black holes considered here, the horizon $r_H$ lies inside the ergoregion, so that $A(r_H)<0$. We therefore divide the original metric by $-A(r)$, obtaining a conformally related metric in which the $tt$ component is fixed to $-1$. On the equatorial
plane we write
\begin{equation}
    \mathrm{d}\hat{s}^2 = -\mathrm{d}t^2
    - \mathcal B(r)^2\,\mathrm{d}r^2
    - \mathcal C(r)^2\,\mathrm{d}\varphi^2 + 2\mathcal D(r)\,\mathrm{d}t\,\mathrm{d}\varphi\,,
\label{conformalmetric}
\end{equation}
where
\begin{align}
    \mathcal B(r)^2 &= -\frac{B(r)}{A(r)}\,, 
    \label{Bconf}\\
    \mathcal C(r)^2 &= -\frac{C(r)}{A(r)}\,,
    \label{Cconf}\\
    \mathcal D(r) &= -\frac{D(r)}{A(r)}\,.
    \label{Dconf}
\end{align}
We are interested in extremal rotating black holes, whose defining
characteristic is the presence of a single, degenerate horizon, where
the outer event horizon and the inner Cauchy horizon coincide. In
Boyer-Lindquist-type coordinates, the horizon is located at a zero of
$g^{rr}$. Since $g^{rr}=-1/B(r)$, extremality implies that $B(r)$ has a quadratic pole at $r=r_H$. In the conformal gauge, we can then write
\begin{equation}
    \mathcal B(r) = \frac{\tilde{\mathcal B}(r)}{r - r_H}\,,
    \quad
    0 < \tilde{\mathcal B}(r_H) < \infty\,.
\label{Bconformal}
\end{equation}
The second near-horizon structure follows from the determinant of the conformal metric. Indeed, from Eq.~\eqref{conformalmetric}, one finds
\begin{equation}
    \det(\hat g_{\mu\nu}) \propto \mathcal B(r)^2\left(\mathcal D(r)^2 - \mathcal C(r)^2\right)\,.
\end{equation}
Requiring the conformal metric determinant to remain finite and nonzero
at the horizon implies that the product
$\mathcal B(r)^2[\mathcal D(r)^2-\mathcal C(r)^2]$ has a finite, nonzero limit as $r\to r_H$. Since $\mathcal B(r)$ has a simple pole at the degenerate horizon, the combination $\mathcal D(r)^2-\mathcal C(r)^2$ must vanish quadratically. Hence, we write
\begin{equation}
    \mathcal D(r)^2 - \mathcal C(r)^2 = \tilde{\Delta}(r)^2 (r - r_H)^2\,,
    \quad
    0 < \tilde{\Delta}(r_H) < \infty\,.
\label{Deltaconformal}
\end{equation}
This isolates the universal near-horizon scaling relevant for the
extremal strong deflection limit, see Sec.~\ref{sec:sdlanalysis}.

We now consider a photon emitted from infinity that approaches the black
hole, reaches a radial turning point at $r_0$, and then escapes to an
observer at infinity. The deflection angle associated with this photon
trajectory is given by the exact integral expression
\begin{equation}
    \hat{\alpha}(r_0) = 2\int_{r_0}^{\infty}\frac{\mathrm{d}\varphi}{\mathrm{d}r}\,\mathrm{d}r - \pi
    \coloneqq \varphi_f(r_0) - \pi\,,
\label{deflangle}
\end{equation}
where \(\varphi_f\) is the total azimuthal shift. Starting from the
conformal metric \eqref{conformalmetric}, one obtains
\begin{equation}
    \frac{\mathrm{d}\varphi}{\mathrm{d}r} = \frac{\mathcal B(r)\left(\mathcal D(r) - J_\pm\right)}{\sqrt{\mathcal D(r)^2-\mathcal C(r)^2}\sqrt{\mathcal C(r)^2 + J_\pm\left(J_\pm - 2\mathcal D(r)\right)}}\,.
\label{integrand1}
\end{equation}
The above expression is readily derived from Eq.~(18) of Ref.~\cite{Bozza2003} after applying the conformal rescaling. Equivalently, using Eqs.~\eqref{Bconformal} and \eqref{Deltaconformal}, we can also write
\begin{equation}
    \frac{\mathrm{d}\varphi}{\mathrm{d}r} = \frac{\tilde{\mathcal B}(r)\left(\mathcal D(r) - J_\pm\right)}{(r - r_H)^2\tilde{\Delta}(r)
    \sqrt{\mathcal C(r)^2 + J_\pm\left(J_\pm - 2\mathcal D(r)\right)}}\,.
\label{integrand1extremal}
\end{equation}
In the above equations, $J_\pm$ denotes the azimuthal angular momentum
per unit energy of the photon. For equatorial trajectories, $J_\pm$
coincides with the photon impact parameter and can be written in terms
of the radial turning point $r_0$ as
\begin{align}
    J_\pm = J_\pm(r_0) &= \mathcal D_0
    \pm \sqrt{\mathcal D_0^2 - \mathcal C_0^2}
    \nonumber\\
    &= \mathcal D_0 \pm (r_0 - r_H)\tilde{\Delta}_0\,.
\label{J}
\end{align}
In Eq.~\eqref{J}, we used a subscript $0$ to denote evaluation at $r_0$.
This expression is obtained by imposing the radial turning-point condition at $r=r_0$, or equivalently by rewriting Eq.~(17) of Ref.~\cite{Bozza2003} in terms of the conformal functions \eqref{Bconf}--\eqref{Dconf}. The two signs in Eq.~\eqref{J} correspond to the two equatorial branches. When a specific rotating solution is considered, we take its spin parameter $a>0$ to describe a black hole rotating in the positive $\varphi$-direction. Then, the minus branch describes photons corotating with the black hole, i.e.~prograde photons, while the plus branch describes counterrotating, i.e.~retrograde, photons. In what follows, we focus on the prograde branch approaching the degenerate horizon and simply write $J \coloneqq J_-$.

We conclude this section by defining $r_m$ as the radial coordinate of the outermost unstable equatorial circular photon orbit of the prograde family. In the conformal gauge, the circular-orbit condition is obtained from Eq.~(31) of Ref.~\cite{Gyulchev2007}, finding
\begin{equation}
    \left(\mathcal C_0^2\right)' - 2J\mathcal D'_0 = 0\,.
\label{unstorbitradius}
\end{equation}
Here and throughout the paper, primes denote differentiation with respect
to the radial coordinate $r$. At the degenerate horizon, Eq.~\eqref{Deltaconformal} implies
\begin{equation}
    \mathcal D_H^2 = \mathcal C_H^2\,,
    \qquad
    J_H = \mathcal D_H\,,
\end{equation}
where the subscript $H$ indicates evaluation at $r_H$. Moreover, differentiating Eq.~\eqref{Deltaconformal} and evaluating the result at $r_H$, one obtains
\begin{equation}
    2\mathcal D_H\mathcal D'_H -
    2\mathcal C_H\mathcal C'_H = 0\,.
\end{equation}
Using these relations, Eq.~\eqref{unstorbitradius} is identically satisfied at $r=r_H$. Thus, an equatorial circular photon orbit is
always present at the degenerate horizon. In what follows, we restrict
to cases with no additional unstable circular orbit outside $r_H$, so
that $r_m=r_H$.

\section{Strong deflection limit analysis}
\label{sec:sdlanalysis}

In this section, we derive the strong-deflection-limit expansion of the deflection angle in Eq.~\eqref{deflangle}, which diverges when the radial turning point $r_0$ approaches the critical circular photon orbit $r_m = r_H$. As already mentioned, photons with $r_0$ sufficiently close to $r_m$ can perform one or more revolutions around the black hole before eventually escaping to infinity. As we shall see, the condition $r_m = r_H$ changes the singular structure of the deflection integral and requires a more careful expansion. We now show how the strong deflection limit can be systematically formulated in this extremal setting.

The first step is to replace the integration variable $r$ in Eq.~\eqref{deflangle} with the new variable
\begin{equation}
    z = 1 - \frac{r_0}{r}\,.
\end{equation}
This change of variable compactifies the integration domain, mapping $r\in[r_0,\infty)$ to $z\in[0,1]$. The deflection angle can then be written as
\begin{equation}
    \hat{\alpha}(r_0) = 2\int_0^1 \frac{\mathrm{d} \varphi}{\mathrm{d}z}\mathrm{d}z - \pi = \varphi_f(r_0) - \pi\,,
\end{equation}
where
\begin{equation}
    \frac{\mathrm{d}\varphi}{\mathrm{d}z} = \frac{r_0\,\tilde{\mathcal B}(r(z))
    \left(\mathcal D(r(z)) - J\right)
    }{\left(\delta + r_H z\right)^2 \tilde{\Delta}(r(z))\sqrt{\mathcal V(r(z);r_0)}}\,.
\label{integrand2}
\end{equation}
Here, we introduced the near-critical parameter
\begin{equation}
    \delta \coloneqq r_0 - r_H \in \mathbb{R}_{>0}\,,
\label{delta}
\end{equation}
so that the strong deflection limit is $\delta \to 0^+$, as well as the quantity
\begin{equation}
    \mathcal V(r(z);r_0) \coloneqq
    \mathcal C(r(z))^2 + J\left[J - 2\mathcal D(r(z))\right]\,.
\label{V}
\end{equation}
Note that all functions without the subscript $0$ are evaluated at
$r(z)=r_0/(1-z)$. In writing Eq.~\eqref{integrand2}, we
have used the relation
\begin{equation}
    r(z) - r_H = \frac{\delta + r_H z}{1 - z}\,.
\end{equation}

So far, no approximations have been made. Eq.~\eqref{integrand2}
displays the two endpoint singularities that control the strong-deflection behavior. The first is the turning-point singularity encoded in $\mathcal V(r(z);r_0)$, already present in the
non-extremal case, while the second is the extremal near-horizon factor
$(\delta+r_H z)^{-2}$. When $r_0\to r_m=r_H$, these two singular
structures overlap at $z=0$, requiring a careful treatment of the expansion.

In analogy with the non-extremal strong deflection limit construction \cite{Bozza2003}, we isolate from the full azimuthal shift $\varphi_f$ a singular contribution obtained from the endpoint expansion of Eq.~\eqref{integrand2}. We can write schematically
\begin{equation}                           \varphi_f(r_0) = \varphi_{f,D}        (r_0) + \varphi_{f,R}(r_0)\,,
\label{sumphiDphiR}
\end{equation}
where $\varphi_{f,D}$ contains the endpoint singularities responsible for the strong-deflection divergence, while $\varphi_{f,R}$ denotes the regular remainder obtained by subtracting this singular approximation from the full integral. In the following, we refer to
$\varphi_{f,D}$ as the singular contribution for brevity, although its
small-$\delta$ expansion also contains finite terms. These finite terms arise from the endpoint integral itself and will contribute to the constant coefficient of the strong-deflection expansion. 

The precise form of $\varphi_{f,D}$ will be presented below. In Sec.~\ref{sec:powerlawdivergence}, we extract the leading contribution to $\varphi_{f,D}$ and show that it gives rise to a power-law divergence. The subleading correction, responsible for the logarithmic divergence, is then derived in Sec.~\ref{sec:logdivergence}. The regular remainder will be discussed later, in Sec.~\ref{sec:regcontrdeflangle}.

\subsection{Power-law divergence}
\label{sec:powerlawdivergence}

As already anticipated above, here we
extract the leading divergent contribution to the total azimuthal shift $\varphi_f$. As shown in Appendix~\ref{app:powerlawdivergence}, keeping the leading non-vanishing terms in the endpoint expansion of Eq.~\eqref{integrand2} gives
\begin{multline}
    \varphi_{f,D}^{(0)}(\delta) = 2r_H
    \frac{\tilde{\mathcal B}_H}{\tilde{\Delta}_H}\,\times
    \\
    \int_0^1\frac{\delta + \mathcal Q_H r_H z}{(\delta + r_H z)^2\sqrt{(\mathcal Q_H - 1)r_H z
    \left[2\delta + (1 + \mathcal Q_H)r_H z\right]}}\mathrm{d}z,
\label{azshiftintegral}
\end{multline}
where we defined
\begin{equation}
    \mathcal Q_H \coloneqq \frac{\mathcal D'_H}{\tilde{\Delta}_H} = \frac{\mathcal C'_H}{\tilde{\Delta}_H}\,.
\label{QH}
\end{equation}
From the integral in Eq.~\eqref{azshiftintegral}, one sees that taking the small-$\delta$ expansion at the level of the integrand is generally not legitimate, because the $\delta\to 0^+$ limit is not uniform over the full integration domain. For this reason, the small-$\delta$ expansion must be carried out on the integrated expression, once the endpoint-sensitive contributions have been properly accounted for. By contrast, factors that remain smooth and well behaved for all $z\in[0,1]$ have been expanded in $\delta$ directly, since their expansion is uniform and does not interfere with the convergence properties of the integral.

Let us then proceed by computing the integral in Eq.~\eqref{azshiftintegral}, finding
\begin{equation}
    \varphi_{f,D}^{(0)}(\delta) =
    2\frac{\tilde{\mathcal B}_H}{\tilde{\Delta}_H}\sqrt{\frac{\mathcal Q_H + 1}{\mathcal Q_H - 1}}
    \frac{\sqrt{1 + \dfrac{2\delta}{(1 + \mathcal Q_H)r_H}}}{\delta\left(1 + \dfrac{\delta}{r_H}\right)}\,.
\label{azimuthalshiftclosedform0}
\end{equation}
Expanding Eq.~\eqref{azimuthalshiftclosedform0} for small $\delta$, we
finally find
\begin{equation}
    \varphi_{f,D}^{(0)}(\delta)
    \simeq 2\frac{\tilde{\mathcal B}_H}{\tilde{\Delta}_H}\sqrt{\frac{\mathcal Q_H + 1}{\mathcal Q_H - 1}}
    \frac{1}{\delta} - \frac{2\tilde{\mathcal B}_H\mathcal Q_H}{r_H\tilde{\Delta}_H\sqrt{\mathcal Q_H^2 - 1}}\,.
\label{azshiftsmalldeltaleading}
\end{equation}
As anticipated, Eq.~\eqref{azshiftsmalldeltaleading} shows that, when the critical prograde orbit lies on the horizon, the leading divergence of the deflection angle is proportional to $1/\delta$. The leading divergence is therefore stronger than that found in the non-extremal strong deflection limit.

\subsection{Logarithmic divergence}
\label{sec:logdivergence}

In this subsection, we derive the subleading logarithmic divergence of
the azimuthal shift. To obtain this term, the leading endpoint
approximation of Sec.~\ref{sec:powerlawdivergence} must be supplemented by the first subleading contribution in the $z\to0$ expansion of the integrand in Eq.~\eqref{integrand2}. At this order, three sources contribute:~the linear correction to the prefactor
$\tilde{\mathcal B}(r(z))/\tilde{\Delta}(r(z))$, the quadratic term in the
expansion of $\mathcal D(r(z))-\mathcal D_H$, and the cubic term in
the endpoint expansion of $\mathcal V(r(z);r_H)$.

Combining these contributions, the leading singular integrand is
multiplied by a factor of the form $1+qz$, where the coefficient $q$
is defined below in terms of the conformal functions and their
derivatives at the horizon. The term proportional to $q$ is precisely
the one responsible for the logarithmic divergence after integration. The details of the expansion are given in
Appendix~\ref{app:logdivergence}. The divergent part of the azimuthal
shift is therefore improved as
\begin{equation}
    \varphi_{f,D}(\delta) \coloneqq
    \varphi_{f,D}^{(0)}(\delta) +
    \varphi_{f,D}^{(1)}(\delta)\,,
\label{azshiftintegral2}
\end{equation}
where $\varphi_{f,D}^{(0)}(\delta)$ is the leading contribution
already derived, while
\begin{multline}
    \varphi_{f,D}^{(1)}(\delta) = 2r_H
    \frac{\tilde{\mathcal B}_H}{\tilde{\Delta}_H}q\,\times
    \\
    \int_0^1\frac{z(\delta + \mathcal Q_H r_H z)}{(\delta + r_H z)^2\sqrt{(\mathcal Q_H - 1)r_H z
    \left[2\delta + (1 + \mathcal Q_H)r_H z\right]}}\mathrm{d}z.
\end{multline}
The coefficient $q$ is given by
\begin{equation}
    q = r_H\left(\frac{\tilde{\mathcal B}'_H}{\tilde{\mathcal B}_H} - \frac{\tilde{\Delta}'_H}{\tilde{\Delta}_H}\right) + \frac{r_H}{2}\left(\frac{\mathcal D''_H}{\mathcal D'_H} - \frac{\mathcal V_3}{\mathcal V_2}\right)\,,
\label{defq}
\end{equation}
with
\begin{equation}
    \mathcal V_2 = \mathcal D_H^{\prime 2} - \tilde{\Delta}_H^2\,,
    \quad
    \mathcal V_3 = \mathcal D'_H\mathcal D''_H - 2\tilde{\Delta}_H\tilde{\Delta}'_H\,.
\label{V2V3}
\end{equation}
The integral defining $\varphi_{f,D}^{(1)}$ can be evaluated
analytically, yielding
\begin{multline}
    \varphi^{(1)}_{f,D}(\delta) =
    \frac{2\tilde{\mathcal B}_H\mathcal Q_H\,q}{r_H\tilde{\Delta}_H\sqrt{\mathcal Q_H^2 - 1}}
    \\
    \times\left[\log\left(
    \frac{\sqrt{1 + \dfrac{2\delta}{(1 + \mathcal Q_H)r_H}} + 1}{\sqrt{1 + \dfrac{2\delta}{(1 + \mathcal Q_H)r_H}} - 1}\right) - \frac{2(\mathcal Q_H - 1)}{\mathcal Q_H}\mathcal S(\delta)\right.
    \\
    \left.-\,\,\frac{1 + \mathcal Q_H}{\mathcal Q_H}\frac{\sqrt{1 + \dfrac{2\delta}{(1 + \mathcal Q_H)r_H}}}{1 + \dfrac{\delta}{r_H}}\right]\,,
\label{phiD1}
\end{multline}
where
\begin{equation}
    \mathcal S(\delta) \coloneqq \sqrt{\frac{\mathcal Q_H + 1}{\mathcal Q_H - 1}}\operatorname{artanh}\left(\sqrt{ \frac{\mathcal Q_H - 1}{1 + \mathcal Q_H + \dfrac{2\delta}{r_H}}}\right)\,.
\label{Sdef}
\end{equation}
Expanding Eq.~\eqref{phiD1} for \(\delta\to0^+\), we then obtain
\begin{multline}
    \varphi^{(1)}_{f,D}(\delta) \simeq
    \frac{2\tilde{\mathcal B}_H\mathcal Q_H q}{r_H\tilde{\Delta}_H\sqrt{\mathcal Q_H^2 - 1}}\bigg[-\log\delta
    \\
    +\,\log\!\left(2(1+\mathcal Q_H)r_H\right) - \frac{2(\mathcal Q_H - 1)}{\mathcal Q_H}\mathcal S(0) -
    \frac{1 + \mathcal Q_H}{\mathcal Q_H}\bigg]\,.
\label{smalldeltalogterm}
\end{multline}
As anticipated, this expression exhibits the logarithmic divergence that
accompanies the leading power-law term.

\subsection{Regular contribution and total deflection angle}
\label{sec:regcontrdeflangle}

We now combine the divergent contributions derived in the previous
subsections and add the regular remainder $\varphi_{f,R}$. Summing
Eqs.~\eqref{azshiftsmalldeltaleading} and \eqref{smalldeltalogterm}, the singular part of the azimuthal shift can
be written as
\begin{equation}
    \varphi_{f,D}(\delta) =
    \frac{\bar{a}_1}{\delta} - \bar{a}_2 \log\delta + \bar{b}_D +
    \mathcal{O}(\delta)\,,
\label{phid_sdl}
\end{equation}
where the coefficients of the power-law and logarithmic divergences are
\begin{align}
    \bar a_1 &\coloneqq 2\frac{\tilde{\mathcal B}_H}{\tilde{\Delta}_H}\sqrt{\frac{\mathcal Q_H + 1}{\mathcal Q_H - 1}}\,,
    \label{a1bar}
    \\
    \bar a_2 &\coloneqq \frac{2\tilde{\mathcal B}_H\mathcal Q_H}{r_H\tilde{\Delta}_H\sqrt{\mathcal Q_H^2 - 1}}q\,,
    \label{a2bar}
\end{align}
respectively, while the finite contribution reads
\begin{multline}
    \bar b_D = \frac{2\tilde{\mathcal B}_H\mathcal Q_H}{r_H\tilde{\Delta}_H\sqrt{\mathcal Q_H^2 - 1}}\bigg\{- 1 + q\bigg[
    \log\!\left(2(1 + \mathcal Q_H)r_H\right)
    \\
    - \frac{2(\mathcal Q_H - 1)}{\mathcal Q_H}\mathcal S(0) -
    \frac{1 + \mathcal Q_H}{\mathcal Q_H}\bigg]\bigg\}\,.
\label{bDbar}
\end{multline}
The regular remainder is defined as the difference between the exact
azimuthal shift, evaluated at $r_0=r_H+\delta$, and the singular endpoint contribution:
\begin{equation}
    \varphi_{f,R}(\delta) = \varphi_f(r_0)\big|_{r_0=r_H+\delta} - \varphi_{f,D}(\delta)\,,
\label{regularpart}
\end{equation}
where $\varphi_{f,D}$ is obtained from Eq.~\eqref{azshiftintegral2}. By construction, the subtraction removes the nonintegrable endpoint behavior as $\delta\to0^+$, so that
$\varphi_{f,R}$ admits a finite limit. Adding the above contribution to Eq.~\eqref{phid_sdl}, we can finally write
\begin{equation}
    \varphi_f(\delta) = \frac{\bar{a}_1}{\delta} - \bar{a}_2 \log(\delta) + \bar{b}_D + \bar{b}_R + \mathcal{O}(\delta)\,,
\end{equation}
where we defined 
\begin{equation}
    \bar{b}_R \coloneqq \lim_{\delta \to 0^+} \varphi_{f,R}(\delta)\,.
\end{equation}
The deflection angle is then obtained simply as
\begin{align}
    \hat{\alpha}(\delta) &= \varphi_f(\delta) - \pi
    \nonumber\\
    &= \frac{\bar{a}_1}{\delta} - \bar{a}_2 \log(\delta) + \bar{b} + \mathcal{O}(\delta)\,,
\label{deflectionanglefinal}
\end{align}
where we also defined
\begin{equation}
    \bar{b} \coloneqq
    \bar{b}_D + \bar{b}_R - \pi\,.
\label{bbar}
\end{equation}
Eqs.~\eqref{a1bar}, \eqref{a2bar}, and \eqref{bbar} define the strong-deflection coefficients in the extremal case.

We note that the integral in Eq.~\eqref{regularpart} is generally the most difficult contribution to evaluate explicitly and must be computed case by case. As we shall see, however, in the extremal Kerr spacetime this term can be obtained analytically.

\subsection{Extremal Kerr black hole:~deflection angle}
\label{sec:deflectionangleKerr}

We now specialize our results to the extremal Kerr spacetime. In
units where $2M=1$, the extremal limit corresponds to $a=1/2$, and
the degenerate horizon is located at $r_H=1/2$ \cite{Bozza2003}. Using Eqs.~\eqref{Bconf}--\eqref{Dconf}, from the equatorial Kerr metric we obtain the rescaled metric coefficients as
\begin{align}
    \mathcal B(r) &= \frac{1}{r - \dfrac12}\sqrt{\frac{r^3}{1 - r}}\,,
    \\
    \mathcal C(r) &= \frac12\sqrt{
    \frac{1 + r + 4r^3}{1 - r}}\,,
    \\
    \mathcal D(r) &= \frac{1}{2(1 - r)}\,,
\end{align}
from which we deduce that
\begin{equation}
    \mathcal D(r)^2-\mathcal C(r)^2 =
    \left(r - \frac12\right)^2
    \frac{r^2}{(1 - r)^2}\,.
\end{equation}
The conformal function $\mathcal B(r)$ has the expected simple pole at
the degenerate horizon, while $\mathcal D(r)^2-\mathcal C(r)^2$ has the corresponding double root. Extremal Kerr therefore belongs to the class of geometries we are considering. By the general argument of Sec.~\ref{sec:prelanddeflangle}, an equatorial circular photon orbit is
present at the degenerate horizon; for prograde photons in extremal
Kerr, this is the relevant orbit governing the strong deflection limit,
so that $r_m=r_H=1/2$.

\begin{figure}[!t]
\includegraphics[width=0.95\linewidth]{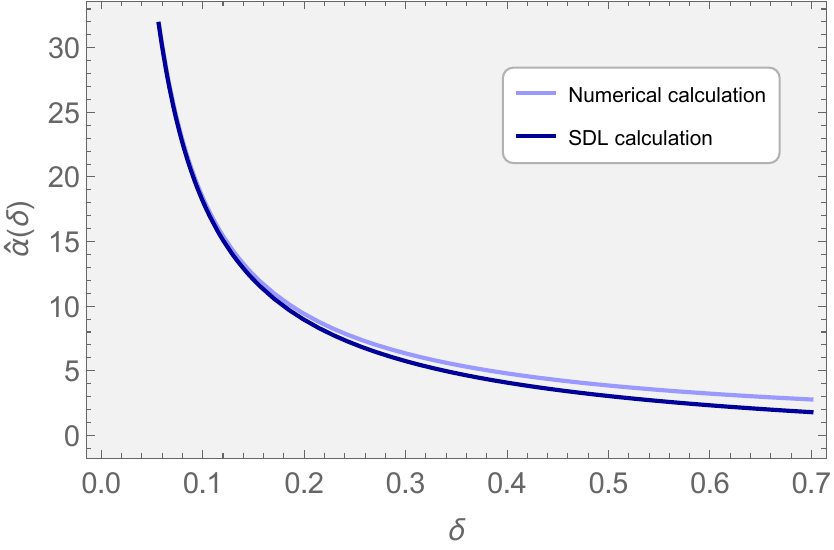}
\caption{Deflection angle for equatorial prograde photon trajectories in the extremal Kerr spacetime, shown as a function of $\delta$. The light blue curve is the result of a numerical calculation, while the dark blue curve is the analytic strong deflection limit approximation, labeled SDL in the legend, given by Eq.~\eqref{deflectionangleKerr}. As expected, for small values of $\delta$, the SDL curve agrees well with the numerical result.}
\label{fig:comparisonSDLnumerical}
\end{figure}

The strong deflection limit coefficients $\bar{a}_1$, $\bar{a}_2$, and $\bar{b}_D$ are obtained by direct substitution of the extremal Kerr metric functions and their derivatives into Eqs.~\eqref{a1bar}, \eqref{a2bar}, and \eqref{bDbar}. This gives
\begin{align}
    &\bar{a}_1 = \sqrt{3}\,,
    \label{a1barKerr}
    \\
    &\bar{a}_2 = \frac{4\sqrt{3}}{9}\,,
    \label{a2barKerr}
    \\
    &\bar b_D = -2\sqrt{3} + \frac{4\sqrt{3}}{9}\log{(3)} - \frac{4}{3}\operatorname{arcoth}\!\big(\sqrt{3}\big)\,.
    \label{bDbarKerr}
\end{align}
As already mentioned, the evaluation of the regular contribution $\bar{b}_R$ is more delicate, since it requires the finite part of the full integral after subtraction of the endpoint singularities. In the extremal Kerr case, this contribution can nevertheless be computed analytically through a matched asymptotic expansion. The details are given in Appendix~\ref{app:maeKerr}; here we only quote the result:
\begin{multline}
    \bar{b}_R = \frac{2}{9}\left(4\,\sqrt{3}\log\left(3 - \sqrt{3}\right)\right.
    \\
    \left.+\,\,6\operatorname{arcoth}\!\left(\sqrt{3} \right) - 6 + 15\,\sqrt{3}\right)\,.
\label{bRKerr}
\end{multline}
Combining the coefficients above, the deflection angle in the extremal Kerr spacetime reads
\begin{multline}
    \hat{\alpha}(\delta) = \frac{\sqrt{3}}{\delta} - \frac{4 \sqrt{3}}{9}\log(\delta) 
    \\
    +\,\frac{4}{9}\left[\sqrt{3}\log\left(18\,\big(2 - \sqrt{3}\big)\right) + 3\,\big(\sqrt{3} - 1\big)\right] - \pi\,.
\label{deflectionangleKerr}
\end{multline}
In Fig.~\ref{fig:comparisonSDLnumerical}, Eq.~\eqref{deflectionangleKerr} is plotted and compared against the numerical calculation, showing excellent agreement for small values of $\delta$.

\section{Higher-order images}
\label{sec:higherorderimages}

We now derive the angular positions of higher-order images from the strong-deflection expansion of Sec.~\ref{sec:sdlanalysis}. In the scenario considered here, where the observer, the source, and the photon trajectory all lie in the equatorial plane, the lens equation can be written as \cite{BozzaScarpetta2007}
\begin{equation}
    \varphi_O-\varphi_S = \hat{\alpha}(\delta) + \pi \mod 2\pi\,.
\end{equation}
Equivalently, we can write
\begin{equation}
    \hat{\alpha}(\delta) = \varphi_O - \varphi_S - \pi + 2\pi n\,,
\label{lensequation1}
\end{equation}
where $n$ is a positive integer counting the number of complete windings performed by the photon around the black hole before reaching the observer. This standard lensing convention should not be confused with classifications based on the number of half-orbits, which are sometimes used to describe higher-order photon rings in black-hole images \cite{Johnson-2020,Broderick-2022,BK-Tsupko-2022}. In Eq.~\eqref{lensequation1}, $\varphi_O$ and $\varphi_S \in [-\pi,\pi]$ are the azimuthal coordinates of the observer and the source, respectively. Fixing the origin of the azimuthal coordinate so that $\varphi_O = \pi$, the lens equation becomes
\begin{equation}
    \hat{\alpha}(\delta)=2\pi n-\varphi_S\,.
\label{lensequation2}
\end{equation}
Note that, in this work, we use this lens equation only to describe the sequence of higher-order images generated by prograde photons. Retrograde photons are instead described by the standard strong deflection limit expansion \cite{Bozza2003}.
 Substituting Eq.~\eqref{deflectionanglefinal} into the lens equation yields
\begin{equation}
    \frac{\bar{a}_1}{\delta} - \bar{a}_2 \log(\delta) + \bar{b} = 2 \pi n - \varphi_S\,.
\label{lenseqalphaexpl}
\end{equation}
Since the logarithmic term is subleading with respect to the power-law term in the limit $\delta \to 0^+$, the solution can be obtained iteratively. At leading order, we obtain
\begin{equation}
    \delta_n^{(0)} = \frac{\bar a_1}{\ell_n}\,,
\label{leadingsolution}
\end{equation}
where we defined
\begin{equation}
    \ell_n \coloneqq 2\pi n - \varphi_S - \bar b\,.
\label{ln}
\end{equation}
Then, evaluating the logarithm in Eq.~\eqref{lenseqalphaexpl} at the leading solution \eqref{leadingsolution}, and considering that $\bigl|\log(\bar{a}_1/\ell_n)\bigr|/\ell_n \ll 1$ for large $\ell_n$, finally gives
\begin{equation}
    \delta_n \simeq \frac{\bar{a}_1}{\ell_n}\left[1 + 
    \frac{\bar{a}_2}{\ell_n}
    \log\left(\frac{\ell_n}{\bar{a}_1}\right)\right]\,.
\label{deltansubleading}
\end{equation}
To convert this result into an angular position, we expand the impact parameter near the critical orbit as
\begin{equation}
    J \simeq J_H + J_{1,H}\,\delta\,, \quad J_{1,H} \coloneqq \left.\frac{\mathrm{d}J}{\mathrm{d}r_0}\right|_{r_0=r_H}\,.
\end{equation}
For an observer in the asymptotic region, $J\simeq \theta D_{OL}$, where $D_{OL}$ is the observer-lens distance, and $\theta$ is the angular separation between the photon’s arrival direction and the direction to the black hole. Therefore, the angular position of the $n$-th higher-order image is
\begin{align}
    \theta_n &\simeq \frac{J_H}{D_{OL}} + \frac{J_{1,H}}{D_{OL}}\,\delta_n 
    \nonumber\\
    &= \frac{J_H}{D_{OL}} + \frac{J_{1,H}}{D_{OL}} \frac{\bar{a}_1}{\ell_n}\left[1 + \frac{\bar{a}_2}{\ell_n}\log\left(\frac{\ell_n}{\bar{a}_1}\right)\right]\,.
\label{thetan}
\end{align}
As one can deduce from Eq.~\eqref{thetan}, the image sequence accumulates at
\begin{equation}
    \theta_\infty \coloneqq \lim_{n \to \infty} \theta_n = \frac{J_H}{D_{OL}}\,,
\end{equation}
where $\theta_\infty$ represents the limiting angular position of the image sequence, corresponding to the shadow edge on the prograde branch. Therefore, the relative angular separation between the $n$-th image and the shadow boundary is expressed as
\begin{equation}
    \frac{s_n}{\theta_\infty} \coloneqq \frac{\theta_n - \theta_\infty}{\theta_\infty} = \frac{\tilde{a}_1}{\ell_n}\left[1 + \frac{\bar{a}_2}{\ell_n}\log\left(\frac{\ell_n}{\bar{a}_1}\right)\right]\,,
\label{relangseparation}
\end{equation}
where we introduced the rescaled strong deflection limit coefficient $\tilde{a}_1$ as
\begin{equation}
    \tilde{a}_1 \coloneqq \frac{J_{1,H}}{J_H}\,\bar{a}_1\,.
\label{rescaleSDLcoeff}
\end{equation}
Unlike in the non-extremal case, where the higher-order images approach the shadow edge exponentially, in the extremal case their separation from the shadow edge decreases as a power law, $s_n\sim 1/n$. This behavior is a direct consequence of the leading power-law divergence of the deflection angle.

For comparison, Eq.~\eqref{lenseqalphaexpl} can also be solved exactly, obtaining
\begin{equation}
    \delta_n = \frac{\bar{a}_1 / \bar{a}_2}{W_0\left(\dfrac{\bar{a}_1}{\bar{a}_2}\exp\left(\dfrac{\ell_n}{\bar{a}_2}\right)\right)}\,,
\end{equation}
where \(W_0\) denotes the principal branch of the Lambert function, and thus
\begin{equation}
    \frac{s_n}{\theta_\infty} = \frac{\tilde{a}_1 / \bar{a}_2}{W_0\left(\dfrac{\bar{a}_1}{\bar{a}_2}\exp\left(\dfrac{\ell_n}{\bar{a}_2}\right)\right)}\,.
\label{relangularseparationexact}
\end{equation}

\subsection{Extremal Kerr black hole:~image positions}

We now specialize the image-position formula to the case of an extremal Kerr spacetime. For the prograde branch, in units $2M = 1$, one finds $J_H = J_{1,H} = 1$.
Therefore, the rescaled strong-deflection coefficient introduced in Eq.~\eqref{rescaleSDLcoeff} reduces to $\tilde a_1 = \bar a_1$.
Using the value of $\bar{a}_1$ given in Eq.~\eqref{a1barKerr}, together with the Kerr coefficients $\bar{a}_2$ and $\bar{b}$ derived in Sec.~\ref{sec:deflectionangleKerr}, Eq.~\eqref{relangseparation} immediately gives the relative angular separation between the $n$-th higher-order image and the shadow edge.

\begin{figure}[!t]
\includegraphics[width=1.0\linewidth]{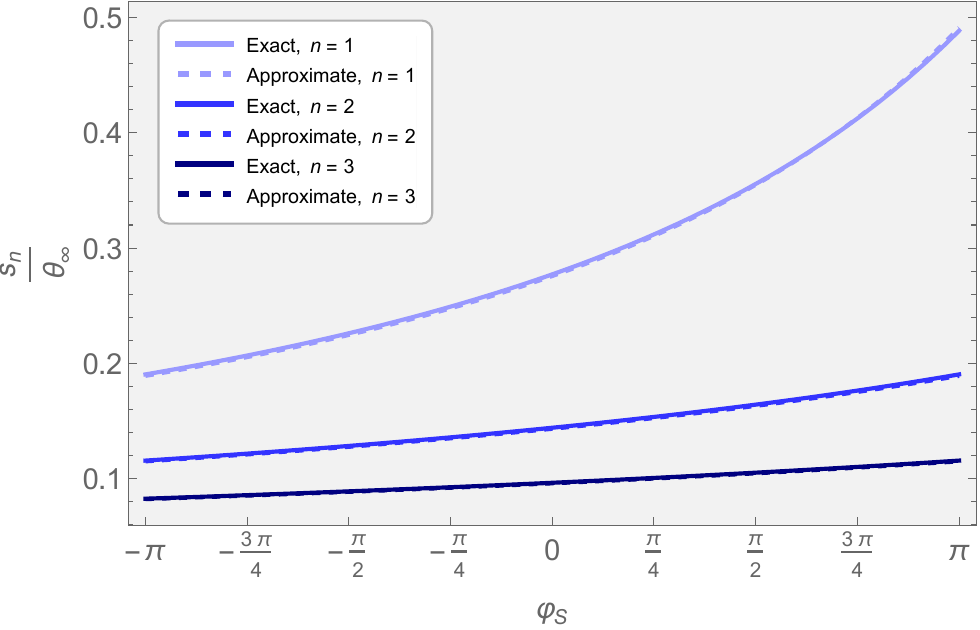}
\centering
\caption{Relative angular position of the first three higher-order images for an extremal Kerr black hole. The horizontal axis is the source azimuth $\varphi_S$, while the vertical axis shows the image angular separation from the shadow edge, normalized by $\theta_\infty$. Each curve corresponds to a certain $n$ (number of loops executed around the black hole before reaching the observer), so that $n = 1$ is the outermost member of the image sequence and larger $n$ accumulate toward $\theta_\infty$. As indicated in the legend, the solid curves represent the approximate results, see Eq.~\eqref{relangseparation}, while the dashed curves represent the exact results, see Eq.~\eqref{relangularseparationexact}.}
\label{fig:imagepositionsKerr}
\end{figure}

The result is shown in Fig.~\ref{fig:imagepositionsKerr} for the first three higher-order images, as a function of $\varphi_S$. For fixed $n$, the separation from the shadow edge increases monotonically with the source azimuth $\varphi_S$. This behavior can be understood from the lens equation:~increasing $\varphi_S$ reduces the total deflection required for a photon with a fixed winding number to connect the source to the observer. In the strong-deflection regime, a smaller deflection corresponds to a trajectory passing slightly farther from the critical orbit, and hence to a larger impact parameter and a larger observed angular separation.

Conversely, increasing $n$ requires the photon to perform an additional winding around the black hole. Such trajectories remain closer to the critical orbit for longer, so their impact parameter approaches the critical value and the corresponding images lie closer to the shadow edge. Accordingly, the $n = 1$ image is the outermost member of the sequence, while images with larger $n$ accumulate toward $\theta_\infty$.

We close this subsection with a brief assessment of the regime of validity of the strong deflection limit approximation we derived. Consider, for instance, Fig.~\ref{fig:imagepositionsKerr}, and focus on the branch $n = 1$. For $\varphi_S = 0$, the plotted values lie in the interval $s_n/\theta_\infty \simeq 0.2\text{--}0.3$, which in the extremal Kerr case corresponds to the same range in $\delta$ (since $s_n/\theta_\infty \simeq (J_{1,H}/J_H)\,\delta$ and $J_{1,H}/J_H=1$ in this case). In this moderately small-$\delta$ regime, however, Fig.~\ref{fig:comparisonSDLnumerical} shows that the difference between the exact deflection angle and its strong deflection limit approximation is already sizable. This behavior can be understood as follows. In the standard (non-extremal) strong deflection limit, the deflection angle has a logarithmic divergence and, as a consequence, the first higher-order image lies exponentially close to the shadow edge. As a result, even the outermost higher-order image is already very close to the critical impact parameter, so the asymptotic description is highly accurate. In the extremal case, instead, the deflection angle features a stronger divergence dominated by an inverse power. For the outermost higher-order image ($n = 1$) this naturally leads to values as large as $\delta = \mathcal{O}(10^{-1})$, which translate into $s_n/\theta_\infty = \mathcal{O}(10^{-1})$ and thus noticeably larger separations from the shadow edge than in the non-extremal case. Since the strong deflection limit is an expansion in small $\delta$, working at $\delta\sim 0.2$--$0.3$ places the first image at the boundary of the asymptotic regime. This motivates the next step:~improving the analytic treatment by extending the deflection-angle expansion to include terms linear in $\delta$.

\section{Improving accuracy:~linear corrections in \texorpdfstring{\(\delta\)}{delta}}
\label{sec:linearcontributions}

Following the discussion at the end of the previous section, here we improve our analytic approximation by extending the deflection-angle expansion beyond the order considered in Sec.~\ref{sec:sdlanalysis}. In the present section, ``improved'' indicates that we retain all terms in the deflection angle up to linear order in $\delta$. In Sec.~\ref{sec:improvdeflangle}, we derive the improved deflection angle, while in Sec.~\ref{sec:improvimpositions} we use it to obtain the corresponding image positions.

\subsection{Improved deflection angle}
\label{sec:improvdeflangle}

We start from the singular contribution $\varphi_{f,D}$. To include
terms up to linear order in $\delta$, the endpoint expansion used in
Secs.~\ref{sec:powerlawdivergence} and \ref{sec:logdivergence} must be
extended by one further order; this amounts to supplementing the correction proportional to $qz$ in Eq.~\eqref{azshiftintegral2}
by a quadratic term. The details of this expansion are given in
Appendix~\ref{app:linear-delta}.

The improved singular contribution takes the form
\begin{equation}
    \varphi_{f,D}(\delta) \coloneqq
    \varphi_{f,D}^{(0)}(\delta)
    +
    \varphi_{f,D}^{(1)}(\delta)
    +
    \varphi_{f,D}^{(2)}(\delta)\,,
\label{azimuthalshiftintegraltot}
\end{equation}
where $\varphi_{f,D}^{(0)}(\delta)$ and
$\varphi_{f,D}^{(1)}(\delta)$ are the integrals already considered in
Secs.~\ref{sec:powerlawdivergence} and \ref{sec:logdivergence},
respectively, while the new contribution is
\begin{multline}
    \varphi_{f,D}^{(2)}(\delta) = 2r_H\frac{\tilde{\mathcal B}_H}{\tilde{\Delta}_H}\,w\,\times
    \\
    \int_0^1\frac{z^2(\delta + \mathcal Q_H r_H z)}{(\delta + r_H z)^2\sqrt{(\mathcal Q_H - 1)r_H z
    \left[2\delta + (1 + \mathcal Q_H)r_H z\right]}}\mathrm{d}z.
\label{azshiftintegral3}
\end{multline}
The coefficient $w$ collects all quadratic corrections in the endpoint
expansion and is defined by
\begin{equation}
    w \coloneqq w_{\mathcal B\Delta} +
    w_{\mathcal D} + w_{\mathcal V} +
    q_{\mathcal B\Delta}\,q_{\mathcal D} + q_{\mathcal B\Delta}\,q_{\mathcal V} + q_{\mathcal D}\,q_{\mathcal V}\,,
\label{wdef}
\end{equation}
where the various subscripts indicate the origin of the corresponding correction:~$\mathcal B\Delta$ refers to the regular prefactor $\tilde{\mathcal B}(r(z))/\tilde{\Delta}(r(z))$, $\mathcal D$ to the numerator
$\mathcal D(r(z))-\mathcal D_H$, and $\mathcal V$ to the function
$\mathcal V(r(z);r_H)^{-1/2}$. The quantities entering $w$ are defined as follows:
\begin{align}
    q_{\mathcal B\Delta} &\coloneqq r_H\left(\frac{\tilde{\mathcal B}'_H}{\tilde{\mathcal B}_H} - \frac{\tilde{\Delta}'_H}{\tilde{\Delta}_H}\right)\,,
    \label{qBDelta}
    \\
    w_{\mathcal B\Delta} &\coloneqq
    q_{\mathcal B\Delta} + \frac{r_H^2}{2}\left[\frac{\left(\tilde{\mathcal B}/\tilde{\Delta}\right)''}{\tilde{\mathcal B}/\tilde{\Delta}}\right]_H\,,
    \label{wBDelta}
    \\
    q_{\mathcal D} &\coloneqq 1 + \frac{r_H\mathcal D''_H}{2\mathcal D'_H}\,,
    \label{qD}
    \\
    w_{\mathcal D} &\coloneqq 1 + \frac{r_H\mathcal D''_H}{\mathcal D'_H} + \frac{r_H^2\mathcal D'''_H}{6\mathcal D'_H}\,,
    \label{wD}
    \\
    q_{\mathcal V} &\coloneqq -\left(
    1 + \frac{r_H\mathcal V_3}{2\mathcal V_2}\right)\,,
    \label{qV}
    \\
    w_{\mathcal V} &\coloneqq r_H^2
    \left(\frac{3\mathcal V_3^2}{8\mathcal V_2^2} - \frac{\mathcal V_4}{2\mathcal V_2}\right)\,.
    \label{wV}
\end{align}
The coefficients $\mathcal V_2$ and $\mathcal V_3$ are given by Eq.~\eqref{V2V3}, while $\mathcal V_4$ is defined as
\begin{equation}
    \mathcal V_4 \coloneqq \frac{\mathcal D_H^{\prime\prime 2}}{4} + \frac{\mathcal D'_H\mathcal D'''_H}{3} - \left(\tilde{\Delta}_H^{\prime 2} + \tilde{\Delta}_H\tilde{\Delta}''_H\right)\,.
\label{V4}
\end{equation}
The quantity $q$ introduced in Eq.~\eqref{defq} is recovered as
\begin{equation}
    q = q_{\mathcal B\Delta} + q_{\mathcal D} + q_{\mathcal V}\,.
\end{equation}

The new integral \eqref{azshiftintegral3} can also be evaluated analytically.
The solution is
\begin{multline}
    \varphi_{f,D}^{(2)}(\delta) = \frac{2\tilde{\mathcal B}_H\mathcal Q_Hw}{r_H\tilde{\Delta}_H\sqrt{\mathcal Q_H^2 - 1}}\left[\sqrt{1 + \frac{2\delta/r_H}{1+\mathcal Q_H}}\right.
    \\
    \left.\hspace{-0.25cm}-\,\,2\frac{2\mathcal Q_H^2 + 2\mathcal Q_H - 1}{\mathcal Q_H(1 + \mathcal Q_H)}\frac{\delta}{r_H}\operatorname{artanh}\left(\frac{1}{\sqrt{1 + \dfrac{2\delta/r_H}{1 + \mathcal Q_H}}}\right)\right.
    \\
    \left.+\,\,\frac{4(\mathcal Q_H - 1)\,\mathcal S(\delta)\,\delta}{\mathcal Q_H\,r_H} + \frac{(1 + \mathcal Q_H)\,\delta}{\mathcal Q_H r_H}\frac{\sqrt{1 + \dfrac{2\delta/r_H}{1 + \mathcal Q_H}}}{1 + \dfrac{\delta}{r_H}}\right].
\label{phi2exact}
\end{multline}
Expanding the above expression up to linear order in $\delta$ yields
\begin{multline}
    \varphi_{f,D}^{(2)}(\delta) \simeq
    \frac{2\tilde{\mathcal B}_H\mathcal Q_Hw}{r_H\tilde{\Delta}_H\sqrt{\mathcal Q_H^2 - 1}}\bigg\{1 + \frac{2\mathcal Q_H^2 + 2\mathcal Q_H - 1}{\mathcal Q_H(1 + \mathcal Q_H)}
    \\
    \times \frac{\delta}{r_H}\log\delta + \frac{\delta}{r_H}\bigg[\frac{\mathcal Q_H^2 + 3\mathcal Q_H + 1}{\mathcal Q_H(1 + \mathcal Q_H)} -
    \frac{2\mathcal Q_H^2 + 2\mathcal Q_H - 1}{\mathcal Q_H(1 + \mathcal Q_H)}
    \\
    \times\,\log\!\left(2(1 + \mathcal Q_H)r_H\right) + \frac{4(\mathcal Q_H - 1)}{\mathcal Q_H}\mathcal S(0)\bigg]
    \bigg\}\,.
\label{azshift3expanded}
\end{multline}

We can now complete the expansion of the singular contribution up to linear order in $\delta$. To this end, we return to Eqs.~\eqref{azimuthalshiftclosedform0} and \eqref{phiD1}, and expand them consistently through $\mathcal{O}(\delta)$. Carrying out these expansions, and then combining with Eq.~\eqref{azshift3expanded}, the singular contribution can be written as
\begin{multline}
    \varphi_{f,D}(\delta) \simeq \frac{\bar{a}_1}{\delta} - \bar{a}_2\log(\delta) + \bar{b}_D 
    \\
    -\,\bar{c}_{1D}\,\delta\log(\delta) + \bar{c}_{2D}\,\delta\,,
\end{multline}
where $\bar{a}_1$ and $\bar{a}_2$ are given in Eqs.~\eqref{a1bar} and \eqref{a2bar}, respectively, while the new coefficients read
\begin{equation}
    \bar c_{1D} = -\frac{2\tilde{\mathcal B}_H\left(2\mathcal Q_H^2 + 2\mathcal Q_H - 1\right)w}{r_H^2\tilde{\Delta}_H(1 + \mathcal Q_H)
    \sqrt{\mathcal Q_H^2 - 1}}\,,
\label{c1D}
\end{equation}
and
\begin{multline}
    \bar c_{2D} = \frac{2\tilde{\mathcal B}_H\mathcal Q_H}{r_H^2\tilde{\Delta}_H\sqrt{\mathcal Q_H^2 - 1}}\bigg\{\frac{2\mathcal Q_H^2 + 2\mathcal Q_H - 1}{2\mathcal Q_H(1 + \mathcal Q_H)}
    \\
    +\,q\frac{2\mathcal Q_H^2 + 2\mathcal Q_H - 1}{\mathcal Q_H(1 + \mathcal Q_H)} + w\bigg[\frac{1}{1 + \mathcal Q_H} + \frac{1 + \mathcal Q_H}{\mathcal Q_H} 
    \\
    \hspace{0.3cm}-\,\frac{2\mathcal Q_H^2 + 2\mathcal Q_H - 1}{\mathcal Q_H(1 + \mathcal Q_H)}\log\!\left(2(1+\mathcal Q_H)r_H\right)
    \\
    + \frac{4(\mathcal Q_H - 1)}{\mathcal Q_H}\mathcal S(0)\bigg]\bigg\}\,.
\label{c2D}
\end{multline}
At this order, the constant term in the singular contribution is also
shifted by the new contribution \eqref{azshiftintegral3}. Explicitly,
we have
\begin{multline}
    \bar b_D = \frac{2\tilde{\mathcal B}_H\mathcal Q_H}{r_H\tilde{\Delta}_H\sqrt{\mathcal Q_H^2 - 1}}\bigg\{w - 1 + q\bigg[
    \log\!\left(2(1 + \mathcal Q_H)r_H\right)
    \\
    - \frac{2(\mathcal Q_H - 1)}{\mathcal Q_H}\mathcal S(0) -
    \frac{1 + \mathcal Q_H}{\mathcal Q_H}\bigg]\bigg\}\,.
\label{bDbarnew}
\end{multline}
Thus, from this point on, $\bar b_D$ denotes the constant coefficient at the improved order.

Finally, we have to include the regular contribution, now understood with the improved singular subtraction in Eq.~\eqref{azimuthalshiftintegraltot}; at linear order in $\delta$, the corresponding regular remainder has the structure
\begin{equation}
    \varphi_{f,R}(\delta) \simeq \bar{b}_R - \bar{c}_{1R}\,\delta\log\delta + \bar{c}_{2R}\,\delta\,.
\label{bRlinearorder}
\end{equation}
The coefficients $\bar{b}_R = \lim_{\delta\to0^+}\varphi_{f,R}(\delta)$, $\bar{c}_{1R}$, and $\bar{c}_{2R}$ must be determined case by case, either analytically or numerically.

Putting together the singular and regular contributions, the improved deflection angle becomes
\begin{equation}
    \hat{\alpha}(\delta) \simeq \frac{\bar{a}_1}{\delta} - \bar{a}_2\log(\delta) - \bar{c}_1\,\delta\log(\delta) + \bar{c}_2\,\delta + \bar{b}\,,
\label{deflangleimproved}
\end{equation}
where we defined
\begin{align}
    \bar{c}_1 &\coloneqq \bar{c}_{1D} + \bar{c}_{1R}\,,
    \\
    \bar{c}_2 &\coloneqq \bar{c}_{2D} + \bar{c}_{2R}\,,
    \\
    \bar{b} &\coloneqq \bar{b}_D + \bar{b}_R - \pi\,.
\end{align}

\subsection{Improved image positions}
\label{sec:improvimpositions}

Following the same strategy adopted in Sec.~\ref{sec:higherorderimages}, the angular separation between the $n$-th higher-order image and the shadow edge can be obtained by solving the lens equation for the small parameter $\delta$. Using the improved deflection angle \eqref{deflangleimproved}, the lens equation can be written as
\begin{equation}
    \frac{\bar{a}_1}{\delta_n} - \bar{a}_2\log(\delta_n) - \bar{c}_1\,\delta_n\log(\delta_n) + \bar{c}_2\,\delta_n = \ell_n\,,
\label{eq:lens-eq-delta}
\end{equation}
with $\ell_n$ defined by Eq.~\eqref{ln}. Solving iteratively for $\delta$ as in Sec.~\ref{sec:higherorderimages} gives
\begin{multline}
    \delta_n \simeq \frac{\bar{a}_1}{\ell_n}
    \left\{1 + \frac{\bar{a}_2}{\ell_n}
    \log\!\left(\frac{\ell_n}{\bar{a}_1}\right)
    \right.
    \\
    \left.+\,\,\frac{1}{\ell_n^2}\left[
    \bar{a}_2^2\left(\log^2\!\left(\frac{\ell_n}{\bar{a}_1}\right) - \log\!\left(\frac{\ell_n}{\bar{a}_1}\right)\right)\right.\right.
    \\
    \left.\left.+\,\,\bar{a}_1\left(\bar{c}_2 + \bar{c}_1\log\!\left(\frac{\ell_n}{\bar a_1}\right)\right)\right]\right\}\,.
\label{deltan_improved}
\end{multline}
We then expand the impact parameter $J$ up to third order in $\delta$:
\begin{equation}
    J \simeq J_H + J_{1,H}\,\delta + \frac{1}{2}J_{2,H}\,\delta^2 + \frac{1}{6}J_{3,H}\,\delta^3\,,
\end{equation}
where
\begin{equation}
    J_{k,H} \coloneqq \frac{\mathrm{d}^kJ}{\mathrm{d}r_0^k}\Big\rvert_{r_0=r_H}\,.
\end{equation}
This finally yields
\begin{multline}
    \frac{s_n}{\theta_\infty} \simeq
    \frac{\tilde{a}_1}{\ell_n} +
    \frac{\tilde{a}_1\bar{a}_2}{\ell_n^2}
    \log\!\left(\frac{\ell_n}{\bar{a}_1}\right)
    + \frac{1}{\ell_n^2}\frac{J_{2,H}}{2J_H}\,\bar{a}_1^2
    \\
    +\,\frac{1}{\ell_n^3}\Bigg[\tilde{a}_1\bar{a}_2^2\left(\log^2\!\left(\frac{\ell_n}{\bar{a}_1}\right) - \log\!\left(\frac{\ell_n}{\bar{a}_1}\right)\right)
    \\
    \quad+\,\frac{J_{2,H}}{J_H}\bar{a}_1^2\bar{a}_2
    \log\!\left(\frac{\ell_n}{\bar{a}_1}\right)
    + \frac{J_{3,H}}{6J_H}\,\bar{a}_1^3
    \\
    +\,\tilde{a}_1\bar{a}_1\left(\bar{c}_2 +
    \bar{c}_1\log\!\left(\frac{\ell_n}{\bar{a}_1}\right)\right)\Bigg]\,.
\label{sn_improved}
\end{multline}

\subsection{Improved extremal Kerr black hole}
\label{sec:improvedkerr}

\begin{figure}[!t]
    \centering
    \begin{tabular}{c}
    \includegraphics[width=0.95\linewidth]{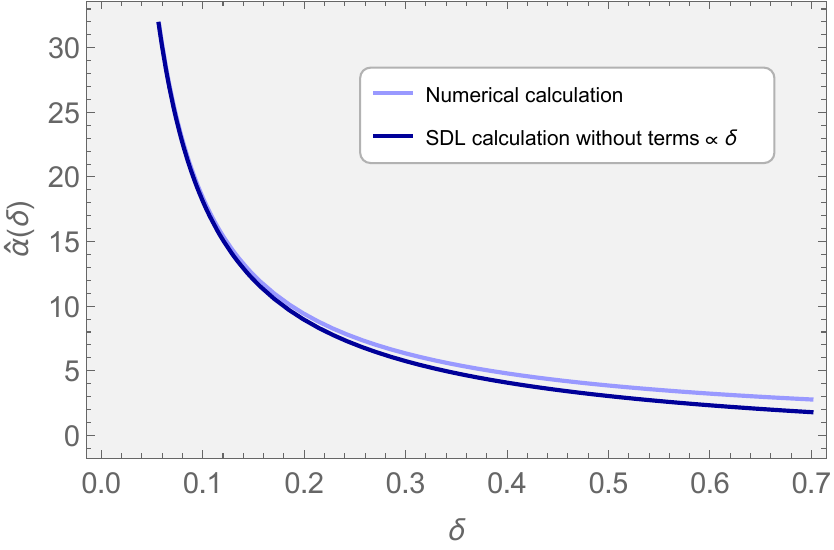}
    \\[0.5em]
    \includegraphics[width=0.95\linewidth]{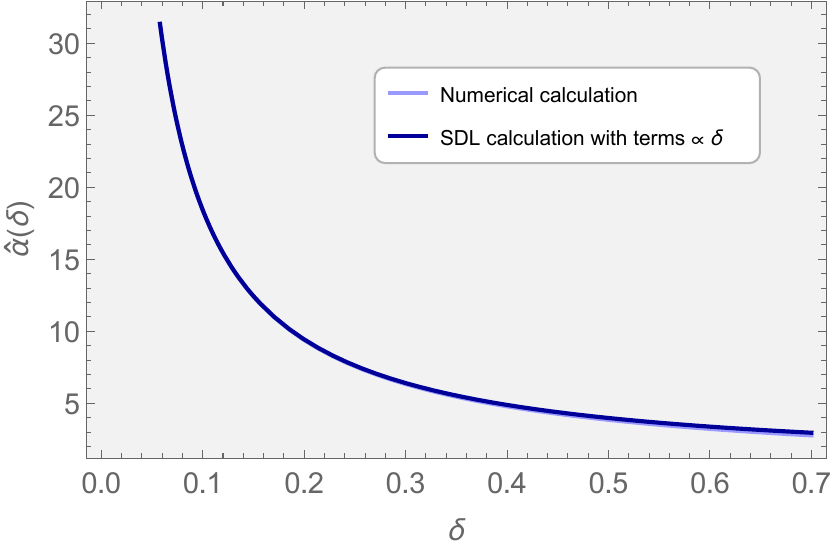}
    \end{tabular}
    \caption{Comparison between the exact deflection angle, computed numerically, and its strong deflection limit approximation (labeled SDL in the legend). We show both the SDL expression retaining only the $1/\delta$, $\log\delta$, and constant terms (dark blue curve, top panel), and the improved SDL including the next-to-leading corrections (dark blue curve, bottom panel).}
    \label{fig:numcalcomparison}
\end{figure}

We now specialize the improved expansion of Sec.~\ref{sec:improvdeflangle} to the extremal Kerr spacetime. The strong deflection limit coefficients $\bar{a}_1$ and $\bar{a}_2$ were already obtained in Eqs.~\eqref{a1barKerr} and \eqref{a2barKerr}. Substituting the extremal Kerr metric functions into Eqs.~\eqref{c1D}, \eqref{c2D}, and \eqref{bDbarnew}, we find
\begin{align}
    \bar{c}_{1D} &= -\frac{22\,\sqrt{3}}{27}\,,
    \\
    \bar{c}_{2D} &= -\bar{c}_{1D}\left(6 - \log3\right) + \frac{8}{3}\operatorname{arcoth}\!\big(\sqrt{3}\big)\,,
    \\
    \bar{b}_D &= -2\sqrt{3} + \frac{4\sqrt{3}}{9}\log{(3)} 
    \nonumber \\
    &\qquad\qquad\qquad\qquad- \frac{4}{3}\operatorname{arcoth}\!\big(\sqrt{3}\big) + \frac{2\,\sqrt{3}}{9}\,,
\end{align}
respectively. Then, we have to determine the regular contribution. Rather remarkably, also at this order the regular remainder can be computed analytically in the extremal Kerr case. Applying the matched asymptotic expansion through linear order in $\delta$, we obtain
\begin{align}
    \bar{b}_R &= \frac{1}{3\sqrt{3}}\left[28 - 4\sqrt{3} + 4\log6\right.
    \nonumber\\
    &\qquad\qquad\qquad\left.+\,\,2\big(2 - \sqrt{3}\big)\log\big(2 - \sqrt{3}\big)\right]\,,
    \\
    \bar{c}_{1R} &= \frac{32}{9\sqrt{3}}\,,
    \\
    \bar{c}_{2R} &= \frac{1}{27\sqrt{3}}\left[12\sqrt{3} - 397 + 30\log2 + 126\log3\right. \nonumber\\
    &\qquad\qquad\qquad\left.+\,\,\big(30 + 36\sqrt{3}\big)\log\big(2 - \sqrt{3}\big)\right]\,.
\end{align}
Combining the singular and regular contributions, we obtain the improved strong-deflection expansion for the extremal Kerr deflection angle as
\begin{multline}
    \hat{\alpha}(\delta) = \frac{\sqrt{3}}{\delta} - \frac{4\sqrt{3}}{9}\log(\delta) + \frac{4}{9}\left[3\big(\sqrt{3} - 1\big)\right.
    \\
    \hspace{-1.45cm}\left.+\,\,\sqrt{3}\log\!\left(18\big(2 - \sqrt{3}\big)\right)\right] - \frac{10\sqrt{3}}{27}\,\delta\log(\delta)
    \\
    +\left[\frac{36 - \sqrt{3}}{81} +\frac{10\sqrt{3}}{27}\log\!\left(18\big(2 - \sqrt{3}\big)\right)\right]\delta - \pi\,.
\label{deflangleKerrimproved}
\end{multline}
In Fig.~\ref{fig:numcalcomparison}, we compare the numerical deflection angle with the analytic approximation. For ease of comparison, the top panel reproduces Fig.~\ref{fig:comparisonSDLnumerical}, where only the $1/\delta$, $\log\delta$, and constant terms are retained. The bottom panel instead shows the improved expression \eqref{deflangleKerrimproved}. The improvement is substantial:~once the linear terms are included, the analytic curve follows the numerical result very closely over the full range of $\delta$ shown.

\section{Applications to other metrics}
\label{sec:KNKS}

Having developed the extremal strong deflection limit expansion in a general form and applied it to the Kerr spacetime, we now turn to other extremal rotating black holes. The purpose of this section is twofold. First, for each metric we identify the region of parameter space in which the equatorial prograde critical orbit coincides with the degenerate horizon, so that the formalism developed above applies. Second, within this regime we compute the corresponding strong deflection limit coefficients and study how the additional parameters of the spacetime affect the deflection angle and the position of the higher-order images.

We begin with the Kerr-Newman spacetime and then consider the Kerr-Sen black hole.

\subsection{Kerr-Newman}
\label{sec:kerrnewman}

Let us now consider an extremal Kerr-Newman black hole
\cite{NewmanJanis19651,NewmanJanis19652,NewmanEtAl1965}. In units where $2M=1$, the
inner and outer horizons are located at \cite{UlbrichtMeinel2015}
\begin{equation}
    r_\pm = \frac{1}{2} \pm \sqrt{\frac{1}{4} - a^2 - Q^2}\,,
\end{equation}
where $Q$ denotes the electric charge. Thus, for $a>0$, extremality requires
\begin{equation}
    a^2 + Q^2 = \frac{1}{4}\,,
\end{equation}
so that the two horizons coincide at $r_+=r_-=r_H=1/2$. Using Eqs.~\eqref{Bconf}--\eqref{Dconf}, from the extremal equatorial Kerr-Newman metric coefficients we find
\begin{align}
    \mathcal B(r) &= \frac{r^2}{\left(r - \frac12\right)\sqrt{
    a^2 - \left(r - \frac12\right)^2}}\,,
    \\
    \mathcal C(r) &= \frac{r}{\sqrt{
    a^2 - \left(r - \frac12\right)^2}}
    \sqrt{r^2 + a^2 + \frac{a^2}{r} - 
    \frac{a^2Q^2}{r^2}}\,,
    \\
    \mathcal D(r) &= \frac{a(r - Q^2)}{
    a^2 - \left(r - \frac12\right)^2}\,,
\end{align}
from which it follows that
\begin{equation}
    \mathcal D(r)^2 - \mathcal C(r)^2 =
    \left(r - \frac12\right)^2\left[ \frac{r^2}{a^2 - \left(r - \frac12\right)^2}\right]^2\,.
\end{equation}
The conformal function $\mathcal B(r)$ has the expected simple pole at
the degenerate horizon, while
$\mathcal D(r)^2-\mathcal C(r)^2$ has the corresponding double root,
as in Eqs.~\eqref{Bconformal} and \eqref{Deltaconformal}.

Restricting to equatorial null geodesics, the circular photon orbits are known in closed form \cite{UlbrichtMeinel2015}; besides the expected horizon root at $r=1/2$, the remaining circular photon orbits are located at
\begin{equation}
    r_{\rm ph}^{\pm} = 1\pm2a\,,
\end{equation}
where, for $a>0$, the upper sign corresponds to the retrograde branch and the lower sign to the prograde branch. The prograde circular photon orbit lies outside the horizon only if
\begin{equation}
    r_{\rm ph}^- > r_H = \frac{1}{2} \quad \Longleftrightarrow \quad
    a < \frac{1}{4}\,.
\end{equation}
Consequently, the critical radius relevant for prograde lensing is
\begin{equation}
    r_m = 
        \begin{cases}
            r_{\rm ph}^- = 1 - 2a\,, & a < \dfrac{1}{4}\,, \\[10pt]
            r_H = \dfrac{1}{2}, & a \ge  \dfrac{1}{4}\,.
        \end{cases}
\label{eq:rm_piecewise_extKN}
\end{equation}
Thus, once $a\ge1/4$, the prograde photon orbit that would otherwise lie outside the black hole is pushed to, or inside, the horizon, and the outermost prograde critical orbit in the exterior coincides with the degenerate horizon. This is precisely the regime to which the extremal strong deflection limit derived in the previous sections applies. Equivalently, using $Q^2=1/4-a^2$, this condition can be written as $|Q| \le \sqrt{3}/4$. There is therefore a sharp threshold along the extremal Kerr-Newman family:~for sufficiently rapid rotation, or equivalently sufficiently small charge, the equatorial prograde critical radius coincides with the horizon. For more weakly rotating, and hence more highly charged, extremal Kerr-Newman black holes, an unstable prograde photon orbit remains outside the horizon and the present extremal expansion does not apply.

\begin{figure*}[!t]
    \centering
    \subfloat[Behavior of $\bar{a}_1$ as a function of the charge $Q$.]{\includegraphics[width=0.46\textwidth,height=0.18\textheight,keepaspectratio]{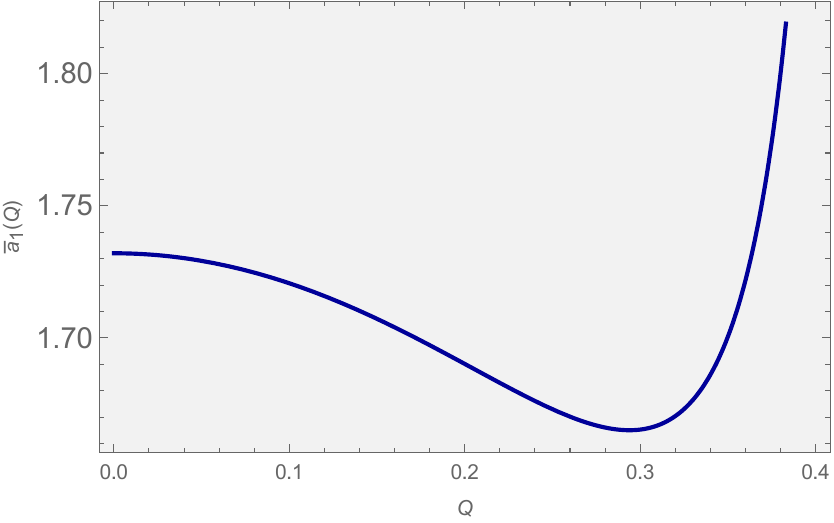}
    \label{fig:abar1}}
    \qquad\qquad\qquad
    \subfloat[Behavior of $\bar{a}_2$ as a function of the charge $Q$.]{\includegraphics[width=0.46\textwidth,height=0.18\textheight,keepaspectratio]{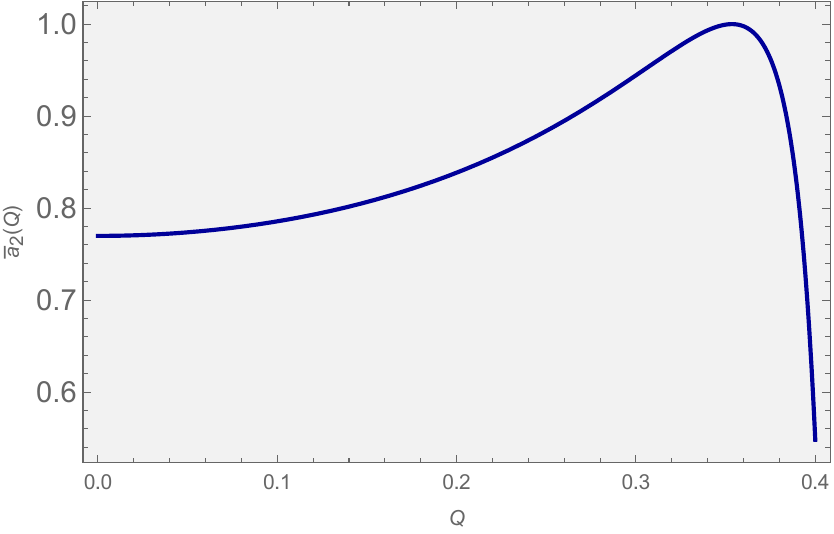}
    \label{fig:abar2}}
    \vspace{0.3em}
    \subfloat[Behavior of $\bar{c}_1$ as a function of the charge $Q$.]{\includegraphics[width=0.46\textwidth,height=0.18\textheight,keepaspectratio]{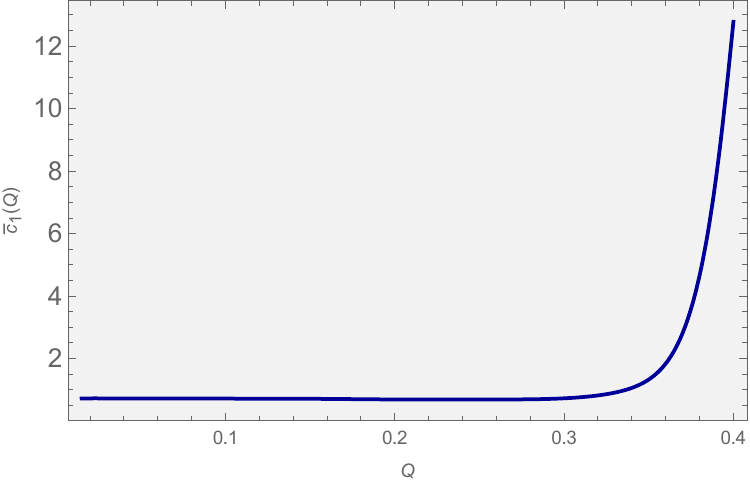}
    \label{fig:cbar1KN}}
    \qquad\qquad\qquad
    \subfloat[Behavior of $\bar{c}_2$ as a function of the charge $Q$.]{\includegraphics[width=0.46\textwidth,height=0.18\textheight,keepaspectratio]{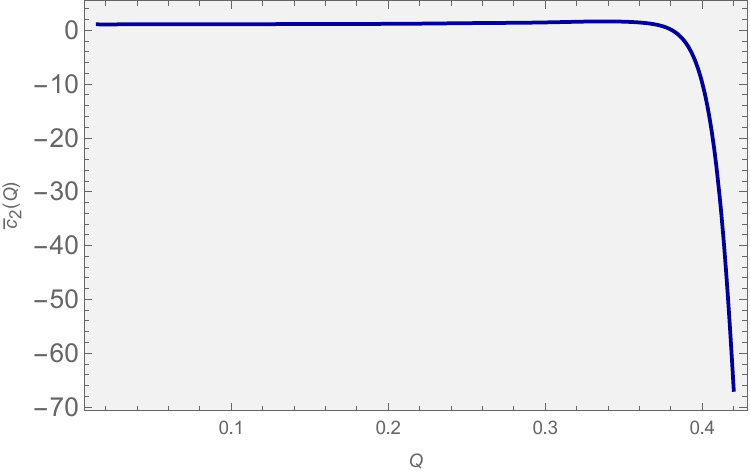}
    \label{fig:cbar2KN}}
    \vspace{0.3em}
    \subfloat[Behavior of $\bar{b}$ as a function of the charge $Q$.]{\includegraphics[width=0.46\textwidth,height=0.18\textheight,keepaspectratio]{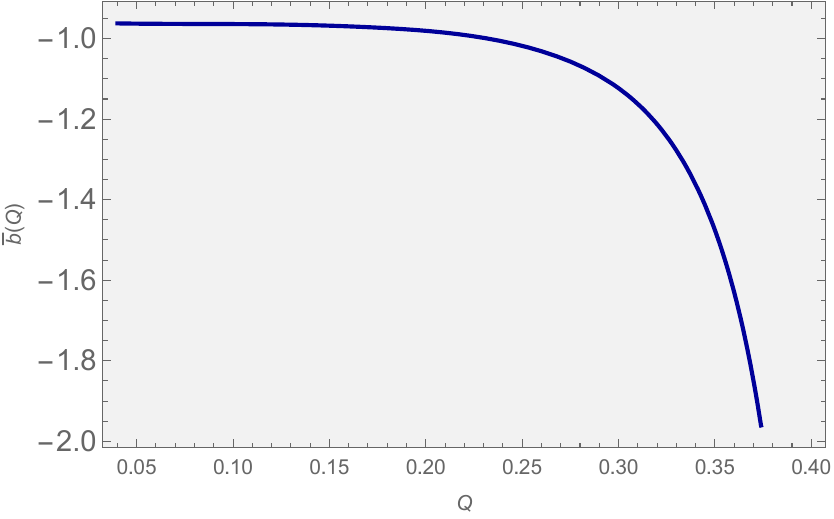}
    \label{fig:bbar}}
    \vspace{0.3cm}
    \caption{Dependence of the strong deflection limit coefficients on the charge $Q$ for an extremal Kerr-Newman black hole.}
    \label{fig:SDLcoefficientsKN}
\end{figure*}

The Kerr-Newman black hole is characterized by two parameters, $a$ and $Q$. However, after imposing extremality, one of them can be eliminated. In what follows, we choose $Q$ as the independent parameter and set
\begin{equation}
    a = \sqrt{\frac{1}{4} - Q^2}\,, \quad |Q| \le \frac{\sqrt{3}}{4} \approx 0.43\,.
\end{equation}
The coefficients controlling the divergent terms in the strong-deflection expansion are obtained by direct substitution of the extremal Kerr-Newman conformal functions into Eqs.~\eqref{a1bar} and \eqref{a2bar}. This gives
\begin{align}
    \bar a_1(Q) &=
    \frac{\sqrt{(1 - 4Q^2)(3 - 16Q^2)}}{2\sqrt{1 - 4Q^2} - 1}\,,
    \\
    \bar a_2(Q) &= \frac{4(1 - 6Q^2)}{(3 - 16Q^2)^{3/2}}\,.
\end{align}
These expressions show explicitly how the power-law and logarithmic divergences depend on the electric charge. The remaining coefficients associated with the endpoint subtraction can also be obtained analytically from the general formulas presented in Sec.~\ref{sec:improvdeflangle}, but their expressions are rather lengthy and not particularly illuminating. The regular coefficients must instead be determined numerically. In Fig.~\ref{fig:SDLcoefficientsKN} we show the behavior of the coefficients as functions of $Q$.

\begin{figure*}[!t]
    \centering
    \begin{minipage}{0.49\textwidth}
        \centering
        \includegraphics[width=\linewidth]{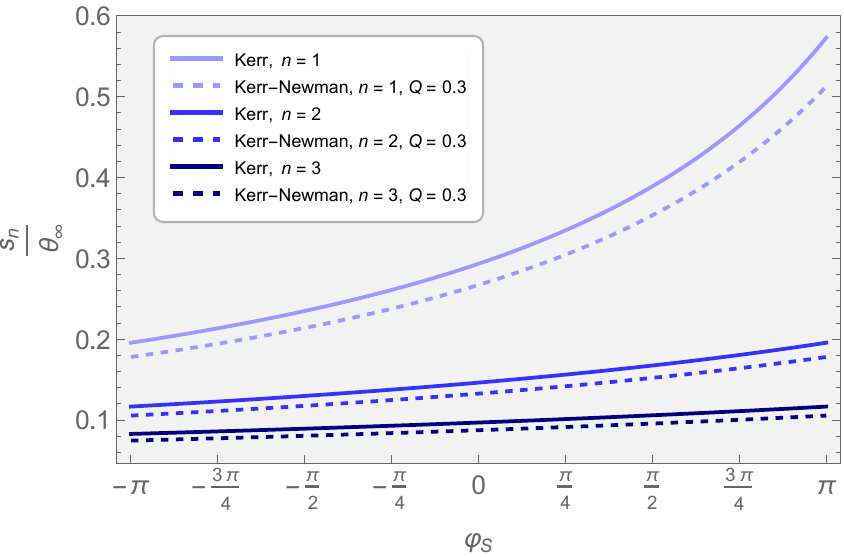}
        \caption{Relative angular position of the first three higher-order images for extremal Kerr and Kerr-Newman black holes. Each curve corresponds to a certain $n$.} \label{fig:improvedimagepositionsKvsKN}
    \end{minipage}
        \hfill
    \begin{minipage}{0.49\textwidth}
        \centering
        \includegraphics[width=\linewidth]{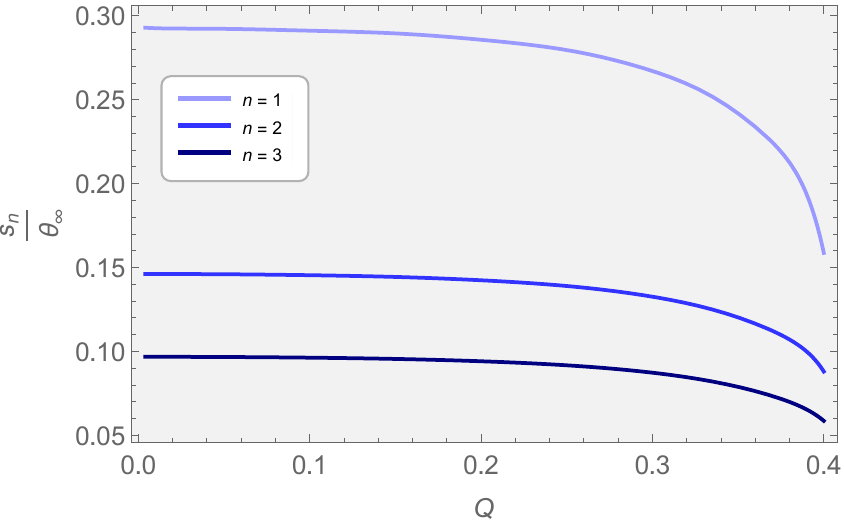}
        \caption{Relative angular position of the first three higher-order images for extremal Kerr-Newman as a function of the charge $Q$. We set $\varphi_S=0$.} \label{fig:improvedimagepositionsKNQ}
    \end{minipage}
\end{figure*}

Finally, Fig.~\ref{fig:improvedimagepositionsKvsKN} shows how the relative angular positions of the first three higher-order images change when going from extremal Kerr to extremal Kerr-Newman, for $Q=0.3$. The Kerr-Newman curves lie systematically below the Kerr curves over the whole range of $\varphi_S$:~at fixed $(n,\varphi_S)$, the images in extremal Kerr-Newman are closer to the shadow edge, corresponding to smaller values of $s_n/\theta_\infty$. The charge-induced shift is largest for $n=1$ and becomes progressively smaller for $n=2$ and $n=3$, consistently with the dominant $1/\ell_n$ scaling in Eq.~\eqref{sn_improved}. In Fig.~\ref{fig:improvedimagepositionsKNQ}, we also show the image positions as functions of $Q$ at fixed source azimuth $\varphi_S=0$. Note that, as $Q$ tends to the critical value $\sqrt{3}/4\approx0.43$ from below, the higher-order images move progressively closer to the shadow edge. This reflects the transition toward the standard logarithmic strong deflection regime, in which the exponential suppression of higher-order image separations is recovered.

\subsection{Kerr-Sen}
\label{sec:kerrsen}

Let us now consider an extremal Kerr-Sen black hole, denoting by $Q$ its electric charge parameter \cite{Sen1992}. In units where $2M=1$, the horizon radii are \cite{Gyulchev2007}
\begin{equation}
    r_{\pm} = \frac{1 - 2Q^2 \pm \sqrt{(1 - 2Q^2)^{2} - 4a^2}}{2}\,.
\end{equation}
For the horizon to exist, we must have
\begin{equation}
    (1 - 2Q^2)^2 - 4a^2 \ge 0\,,
\end{equation}
which in particular implies $Q^{2}\le1/2$. For $a > 0$, the extremality condition is
\begin{equation}
    (1 - 2Q^2)^2 - 4a^{2} = 0\,,
\end{equation}
from which it immediately follows that
\begin{equation}
    a = \frac{1}{2} - Q^2\,.
\end{equation}
The degenerate horizon is then located at
\begin{equation}
    r_{+} = r_{-} = r_H = \frac{1}{2}  - Q^2\,.
\end{equation}
On the equatorial plane, for the extremal Kerr-Sen metric coefficients \cite{Gyulchev2007,Sen1992}, Eqs.~\eqref{Bconf}--\eqref{Dconf} give
\begin{align}
    \mathcal B(r)  &= \frac{r + 2Q^2}{
    r - r_H}\sqrt{\frac{r}{1 - r - 2Q^2}}\,,
    \label{BKS}
    \\
    \mathcal C(r) &= \sqrt{\frac{\left(r + 2Q^2\right)
    \left[r\left(r + 2Q^2\right) + a^2 +
    \dfrac{a^2}{r + 2Q^2}\right]}{1 - r - 2Q^2}}\,,
    \\
    \mathcal D(r) &= \frac{a}{1 - r - 2Q^2}\,.
\end{align}
We then find
\begin{equation}
    \mathcal D(r)^2 - \mathcal C(r)^2 = 
    \left(r - r_H\right)^2\left[\frac{
    r + 2Q^2}{1 - r - 2Q^2}\right]^2\,.
\label{DCKS}
\end{equation}
Inspecting Eqs.~\eqref{BKS} and \eqref{DCKS}, we can conclude that Kerr-Sen belongs to the class of geometries considered in Sec.~\ref{sec:prelanddeflangle}.

The equatorial circular photon
orbits in Kerr-Sen can be obtained from Eq.~\eqref{unstorbitradius} after imposing the extremality condition; besides the horizon
circular orbit, expected from the general argument of
Sec.~\ref{sec:prelanddeflangle}, the remaining solutions are
\begin{equation}
    r_{1,2} = 1 - 2Q^2 \pm \sqrt{1 - 2Q^2}\,.
\end{equation}
Whenever the plus root corresponds to an exterior orbit, it belongs
to the retrograde branch. No additional exterior prograde root appears:~the prograde equatorial circular photon orbit coincides with the degenerate horizon. Therefore, throughout the extremal Kerr-Sen family, the prograde critical radius is $r_m=r_H$.

Unlike in the extremal Kerr-Newman case, there is no additional spin/charge threshold:~once the Kerr-Sen black hole is extremal, the equatorial prograde critical orbit coincides with the horizon. In other words, Kerr-Newman extremality does not by itself force the prograde circular photon orbit to lie on the horizon, whereas Kerr-Sen extremality does.

As in the extremal Kerr-Newman case, the extremality condition allows us to eliminate one of the two parameters. We may either express the lensing quantities in terms of $Q$, by substituting $a=1/2-Q^2$ into the metric coefficients, or equivalently in terms of $a$, using $Q^2=1/2-a$. In what follows, we choose $Q$ as the independent parameter.

\begin{figure*}[!t]
    \centering
    \subfloat[Behavior of $\bar{a}_1$ as a function of the charge $Q$.]{  \includegraphics[width=0.46\textwidth,height=0.18\textheight,keepaspectratio]{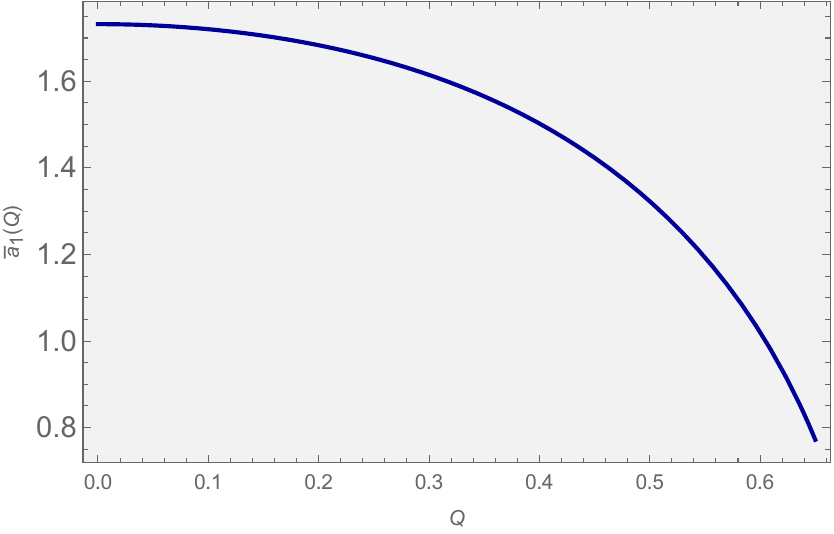}
    \label{fig:abar1KS}}
    \qquad\qquad\qquad
    \subfloat[Behavior of $\bar{a}_2$ as a function of the charge $Q$.]{\includegraphics[width=0.46\textwidth,height=0.18\textheight,keepaspectratio]{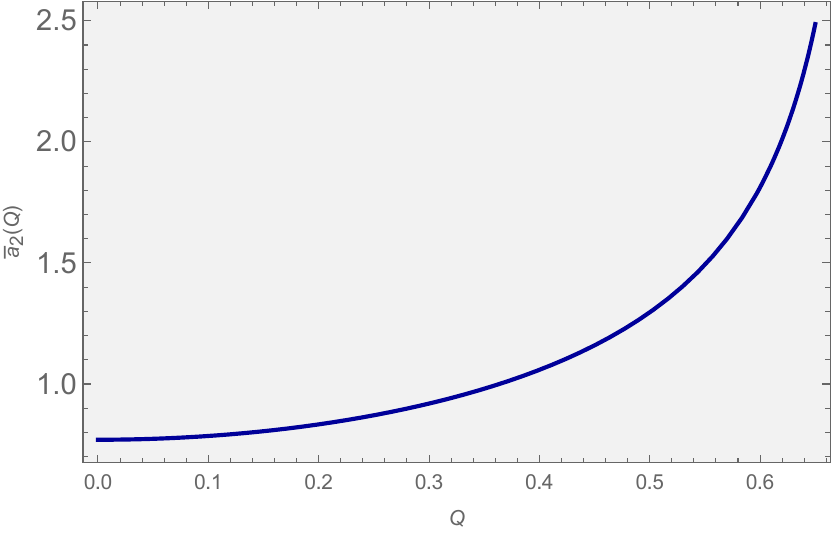}
    \label{fig:abar2KS}}
    \vspace{0.3em}
    \subfloat[Behavior of $\bar{c}_1$ as a function of the charge $Q$.]{\includegraphics[width=0.46\textwidth,height=0.18\textheight,keepaspectratio]{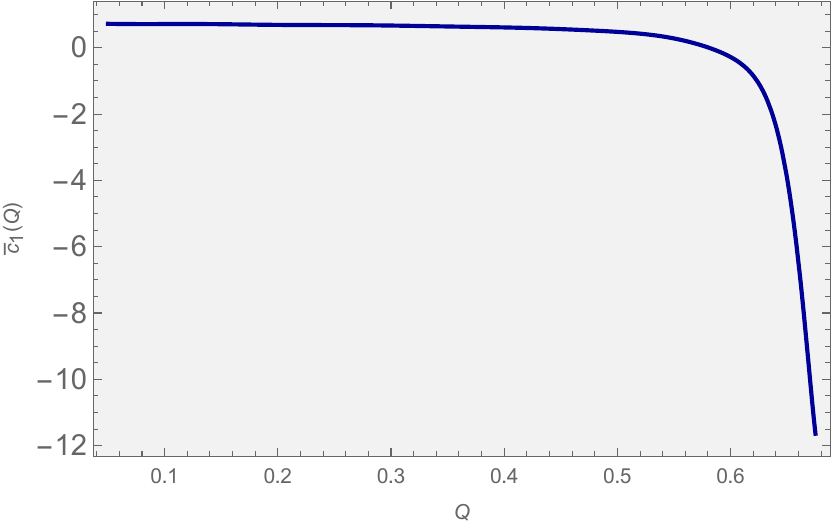}
    \label{fig:cbar1KS}}
    \qquad\qquad\qquad
    \subfloat[Behavior of $\bar{c}_2$ as a function of the charge $Q$.]{\includegraphics[width=0.46\textwidth,height=0.18\textheight,keepaspectratio]{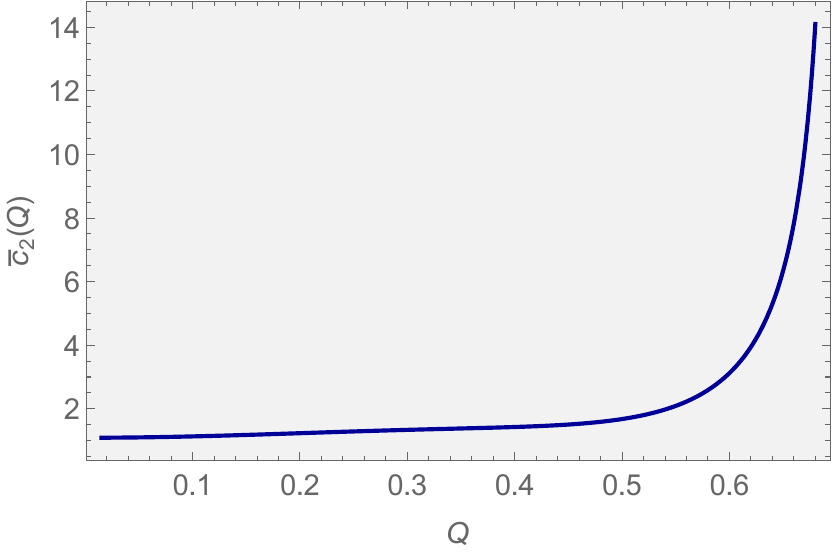}
    \label{fig:cbar2KS}}
    \vspace{0.3em}
    \subfloat[Behavior of $\bar{b}$ as a function of the charge $Q$.]{\includegraphics[width=0.46\textwidth,height=0.18\textheight,keepaspectratio]{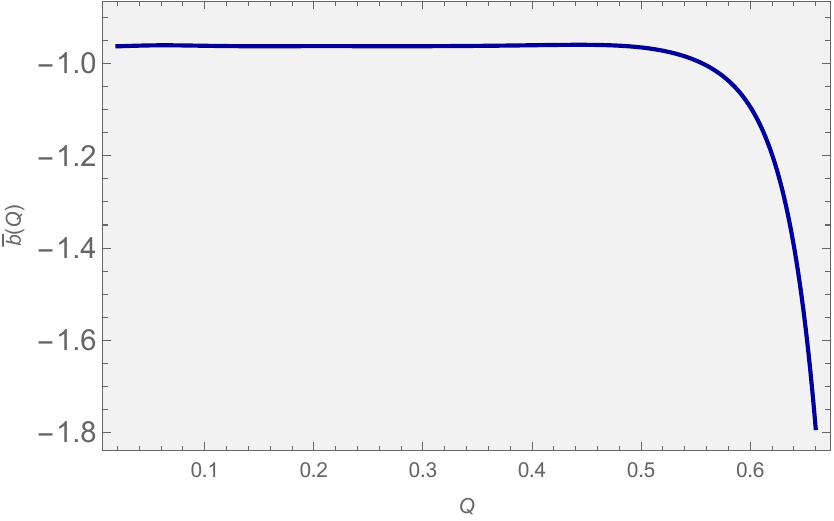}
    \label{fig:bbarKS}}
    \vspace{0.3cm}
    \caption{Dependence of the strong deflection limit coefficients on the charge $Q$ for an extremal Kerr-Sen black hole.}
    \label{fig:SDLcoefficientsKS}
\end{figure*}

\begin{figure*}[!t]
    \centering
    \begin{minipage}{0.48\textwidth}
        \centering
        \includegraphics[width=\linewidth]{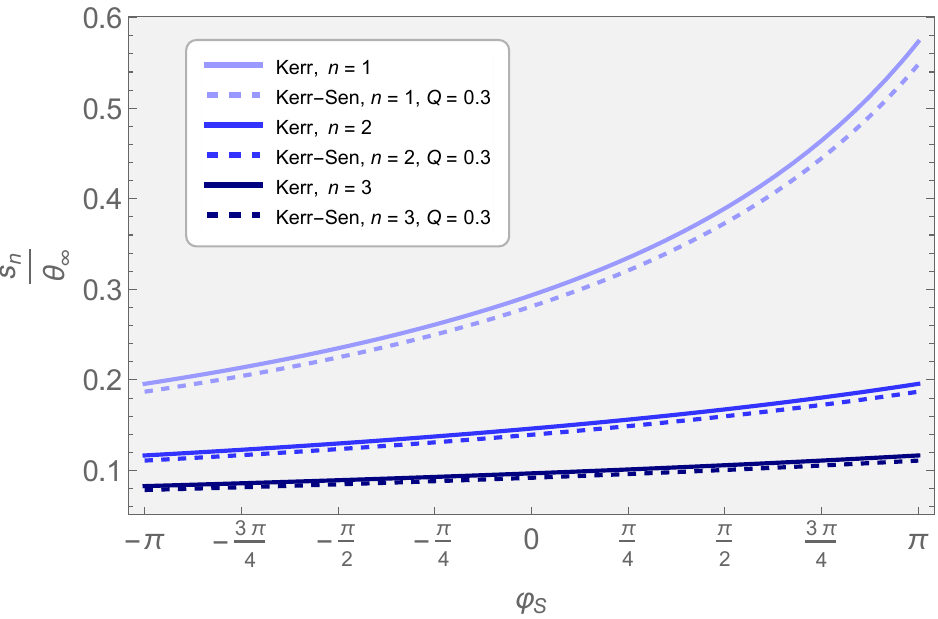}
        \caption{Relative angular position of the first three higher-order images for extremal Kerr and Kerr-Sen black holes. Each curve corresponds to a certain $n$.} \label{fig:improvedimagepositionsKvsKS}
    \end{minipage}
        \hfill
    \begin{minipage}{0.48\textwidth}
        \centering
        \includegraphics[width=\linewidth]{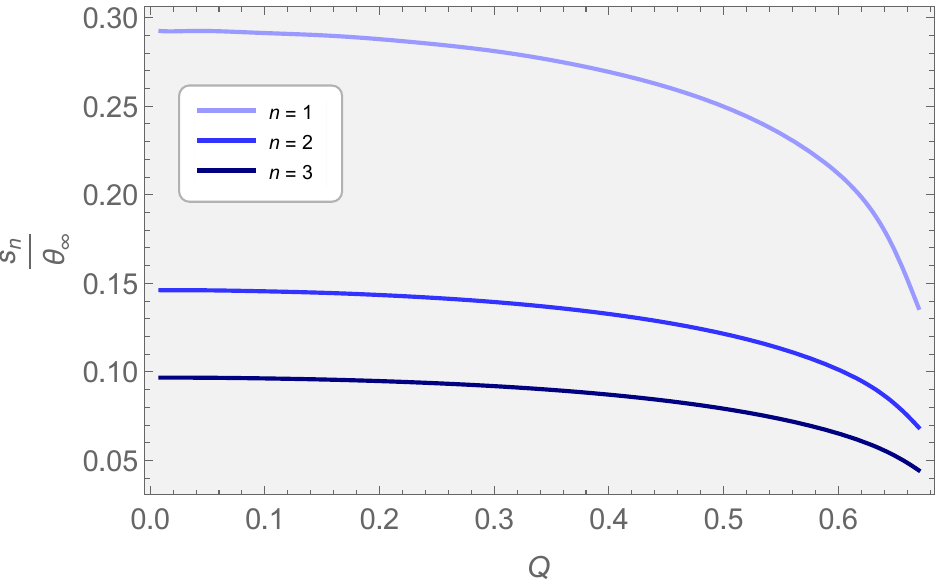}
        \caption{Relative angular position of the first three higher-order images for extremal Kerr-Sen as a function of the charge $Q$. We set $\varphi_S=0$.} \label{fig:improvedimagepositionsKSQ}
    \end{minipage}
\end{figure*}

With the extremality condition imposed, the general expressions of Sec.~\ref{sec:sdlanalysis} can be specialized to the Kerr-Sen metric; the resulting behavior of the strong deflection limit coefficients as a function of the charge is shown in Fig.~\ref{fig:SDLcoefficientsKS}. As in the extremal Kerr-Newman case, the coefficients controlling the two divergent terms take a compact
form:
\begin{align}
    \bar a_1(Q) &= \sqrt{\left(1 - 2Q^2\right)\left(3 + 2Q^2\right)}\,,
    \\
    \bar a_2(Q) &= \frac{4\left(1 + 2Q^2\right)}{\left(3 + 2Q^2\right)\sqrt{\left(1 - 2Q^2\right)\left(3 + 2Q^2\right)}}\,.
\end{align}
The above formulas clearly illustrate how the $1/\delta$ and $\log\delta$ divergences in the deflection angle depend on the Kerr-Sen charge.

Finally, Fig.~\ref{fig:improvedimagepositionsKvsKS} shows how the relative angular positions of the first three higher-order images ($n=1,2,3$) change from extremal Kerr to extremal Kerr-Sen, for $Q=0.3$. The Kerr-Sen curves lie systematically below the Kerr ones across the whole $\varphi_S$ range. In Fig.~\ref{fig:improvedimagepositionsKSQ}, we also plot the position of the images as a function of $Q$ at fixed $\varphi_S=0$.

\section{Quasi-equatorial motion}
\label{sec:smalldeclinations}

Up to this point, we have restricted photon motion to the equatorial plane, $\vartheta=\pi/2$, so that the relevant geometry was described by the $(t,r,\varphi)$ sector of the metric. In order to compute the magnification of higher-order images, however, one must go beyond this one-dimensional lens map and allow for small deviations from the equatorial plane. We therefore restore the polar part of the stationary, axisymmetric line element in Boyer--Lindquist-type coordinates $(t,r,\vartheta,\varphi)$. In the conformal gauge, we write
\begin{multline}
    \mathrm{d}\hat{s}^2 = -\mathrm{d}t^2 - \mathcal B(r,\vartheta)^2\,\mathrm{d}r^2
    - \mathcal F(r,\vartheta)^2\,\mathrm{d}\vartheta^2
    \\
    - \mathcal C(r,\vartheta)^2\,\mathrm{d}\varphi^2
    + 2\mathcal D(r,\vartheta)\,\mathrm{d}t\,\mathrm{d}\varphi\,.
\label{conformalmetricfull}
\end{multline}
Here, in analogy with Eqs.~\eqref{Bconf}--\eqref{Dconf}, we have introduced the quantity
\begin{equation}
    \mathcal F(r,\vartheta)^2 = -
    \frac{F(r,\vartheta)}{A(r,\vartheta)}\,,
\label{Fconf}
\end{equation}
where $F(r,\vartheta)$ denotes the polar metric coefficient in the
original line element, before the conformal rescaling. On the equatorial plane, the metric functions $\mathcal B,\mathcal C,\mathcal D$ reduce to those introduced in
Sec.~\ref{sec:prelanddeflangle}. In what follows, whenever no explicit $\vartheta$-dependence is displayed, all functions are evaluated on the equatorial plane.

\begin{figure}[!t]
\includegraphics[width=0.95\linewidth]{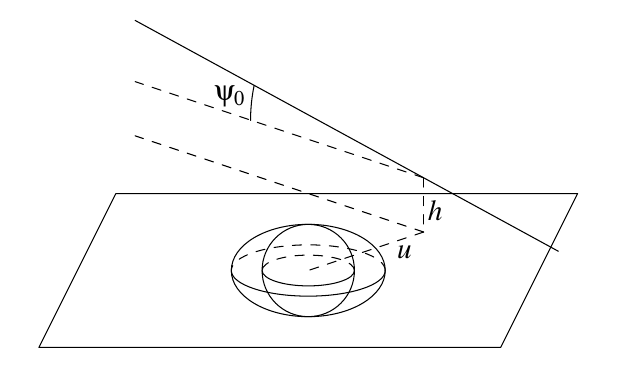}
\caption{Geometrical parametrization of an incoming quasi-equatorial photon trajectory in the asymptotically flat region. The parameter $u$ is the minimum distance between the black hole and the projection of the asymptotic trajectory onto the equatorial plane. The quantity $h$ is the height of the photon above or below the equatorial plane at that point, while $\Psi_0$ is the inclination angle of the asymptote with respect to the equatorial plane. Reproduced from Ref.~\cite{Bozza2003}.}
\label{fig:qelensing}
\end{figure}

The next step is to describe the polar motion. For the class of
spacetimes considered here, we assume Hamilton--Jacobi separability and, more specifically, a Kerr-type angular sector. As in the standard
quasi-equatorial treatment, we denote by $J\coloneqq L/E$ the azimuthal angular momentum per unit energy. Away from the equatorial plane, however, $J$ no longer coincides exactly with the equatorial impact parameter. Introducing also the Carter constant per unit energy squared,
$\mathcal K$, the polar first integral can be written as \cite{Chandrasekhar1983,Bozza2003}
\begin{equation}
    \dot{\vartheta} = \pm
    \frac{E}{\mathcal F(r,\vartheta)^2}\sqrt{\mathcal K
    + a^2\cos^2\vartheta -
    J^2\cot^2\vartheta}\,.
\label{thetadot}
\end{equation}
To describe trajectories displaced from the equatorial plane, we adopt the geometrical parametrization shown in Fig.~\ref{fig:qelensing}, reproduced from Ref.~\cite{Bozza2003}. In the asymptotically flat region, each incoming photon can be associated with the straight line it would follow in the absence of the gravitational field. This asymptotic line is specified by three quantities:~the projected impact parameter $u$, the height $h$ at the point of closest projected approach to the black hole, and the inclination angle $\Psi_0$ with respect to the equatorial plane. These quantities determine the conserved parameters of the geodesic as \cite{Bozza2003}
\begin{align}
    J &= u \cos(\Psi_0)\,, 
    \label{qeJ} \\
    \mathcal K &= h^2 \cos^2(\Psi_0) + (u^2 - a^2) \sin^2(\Psi_0)\,. 
    \label{Cbar}
\end{align}
Since $J$ and $\mathcal K$ are constants of motion, the asymptotic data $(u,h,\Psi_0)$ provide a convenient geometrical way to label the corresponding deflected geodesic.

We now restrict our attention to trajectories that remain close to the equatorial plane. To this end, we introduce the declination $\Psi\coloneqq \pi/2-\vartheta$, so that $\Psi=0$ corresponds to motion on the equatorial plane, while $|\Psi|\ll 1$ describes quasi-equatorial trajectories. The quantity $\Psi$ is a dynamical variable along the geodesic. On the incoming branch, its asymptotic value coincides with the initial inclination angle $\Psi_0$ we introduced before. In the quasi-equatorial regime, i.e., when $|\Psi_0|\ll 1$, Eqs.~\eqref{qeJ} and \eqref{Cbar} reduce to
\begin{align}
    J &\simeq u\,, 
    \\
    \mathcal K &\simeq h^2 + \bar{u}^2\,\Psi_0^2\,, \quad \bar{u} \equiv \sqrt{u^2 - a^2}\,.
\end{align}

In order to follow the evolution of the declination $\Psi$ as the
photon winds around the black hole, we take the ratio between the polar first integral and the azimuthal first integral. The latter follows from the conserved energy and angular momentum associated with the Killing fields $\partial_t$ and $\partial_\varphi$ \cite{Chandrasekhar1983,Bozza2003}. In the conformal gauge, it reads
\begin{equation}
    \dot{\varphi} = E \frac{\mathcal D(r,\vartheta) - J}{\mathcal D(r,\vartheta)^2 - \mathcal C(r,\vartheta)^2
    }\,.
\label{phidotconf}
\end{equation}
Taking the ratio of Eqs.~\eqref{thetadot} and
\eqref{phidotconf}, and working at leading order in the quasi-equatorial expansion, we can then write \cite{Bozza2003}
\begin{equation}
    \frac{\mathrm{d}\Psi}{\mathrm{d}\varphi} = \pm \omega(r(\varphi))\sqrt{\bar\Psi^2 - \Psi^2}\,,
\label{Psievolutionequation}
\end{equation}
where
\begin{equation}
    \omega(r(\varphi)) \coloneqq \frac{\bar u}{\mathcal F(r(\varphi))^2}\frac{
    \mathcal D(r(\varphi))^2 - \mathcal C(r(\varphi))^2}{\mathcal D(r(\varphi)) - J}\,,
\label{omegaconf}
\end{equation}
and
\begin{equation}
     \bar\Psi \coloneqq \sqrt{
    \frac{h^2}{\bar u^2} + \Psi_0^2}\,.
\end{equation}
Eq.~\eqref{Psievolutionequation} has a very simple interpretation:~in the
quasi-equatorial regime, the photon oscillates above and below the equatorial plane with amplitude $\bar\Psi$, while $\omega(r(\varphi))$ plays
the role of an angular frequency for this polar oscillation.

The solution of Eq.~\eqref{Psievolutionequation} can be written as \cite{Bozza2003}
\begin{align}
    &\Psi(\varphi) = \bar{\Psi} \cos(\bar{\varphi} + \varphi_0)\,,
    \\
    &\bar{\varphi} \coloneqq \int_0^\varphi \omega(r(\varphi^\prime))\,\mathrm{d}\varphi^\prime\,,
\end{align}
where $\varphi_0$ is fixed by the initial conditions. The quantity relevant for lensing is the total polar phase accumulated along the trajectory,
\begin{equation}
    \bar{\varphi}_f =
    \int_0^{\varphi_f}
    \omega(r(\varphi'))\,\mathrm{d}\varphi' =
    2\int_{r_0}^{\infty}
    \omega(r)\frac{\mathrm{d} \varphi}{\mathrm{d}r}\,\mathrm{d}r\,,
\label{psibarf}
\end{equation}
where $\varphi_f$ is the total azimuthal shift computed previously in Sec.~\ref{sec:sdlanalysis}.

Using Eqs.~\eqref{Bconformal}, \eqref{Deltaconformal}, and \eqref{integrand1}, the product entering Eq.~\eqref{psibarf}
becomes
\begin{equation}
    \omega(r) \frac{\mathrm{d}\varphi}{\mathrm{d}r} = \bar u
    \frac{\tilde{\mathcal B}(r)\tilde{\Delta}(r)}{\mathcal F(r)^2\sqrt{\mathcal V(r;r_0)}}\,.
\label{omegadphidr}
\end{equation}
Notice that the explicit near-horizon factor responsible for the
power-law divergence of the azimuthal shift cancels in the product $\omega\,\mathrm{d}\varphi/\mathrm{d}r$. The singular structure of the polar phase is
therefore governed only by the factor $\mathcal V(r;r_0)$. Thus, unlike the azimuthal shift, the polar phase has the same type of square-root endpoint singularity as in the non-extremal strong deflection limit, in agreement with the results of Ref.~\cite{Gralla2019}.

Introducing the same compact variable used in Sec.~\ref{sec:sdlanalysis},
\begin{equation}
    z = 1 - \frac{r_0}{r}\,,
\end{equation}
we can rewrite Eq.~\eqref{psibarf} as
\begin{equation}
    \bar\varphi_f = \int_0^1
    \mathcal R_\omega(z;r_0) f_\omega(z;r_0)\,\mathrm{d}z\,,
\label{polarphasez}
\end{equation}
where
\begin{equation}
    \mathcal R_\omega(z;r_0) \coloneqq 2\bar u \frac{r_0}{(1 - z)^2}\frac{\tilde{\mathcal B}(r(z))\tilde{\Delta}(r(z))}{
    \mathcal F(r(z))^2}\,,
\label{Romega}
\end{equation}
and
\begin{equation}
    f_\omega(z;r_0) \coloneqq
    \frac{1}{\sqrt{\mathcal V(r(z);r_0)}}\,.
    \label{fomega}
\end{equation}

By applying the strong deflection limit analysis to
Eq.~\eqref{polarphasez} at leading order, we obtain
\begin{equation}
    \bar\varphi_f(\delta) = -\hat a\log\delta + \hat b\,,
\label{polarphase}
\end{equation}
where the coefficient of the logarithmic term is
\begin{equation}
    \hat a \coloneqq \frac{
    \mathcal R_\omega(0;r_H)}{
    \tilde{\Delta}_Hr_H\sqrt{\mathcal Q_H^2 - 1}}\,,
\label{ahat}
\end{equation}
while the constant contribution is written as
\begin{equation}
    \hat b \coloneqq
    \hat b_D + \hat b_R\,,
\end{equation}
with
\begin{equation}
    \hat b_D \coloneqq \hat a \log\left[
    2(1 + \mathcal Q_H)r_H\right]\,,
\label{bhatD}
\end{equation}
and
\begin{multline}
    \hat b_R \coloneqq
    \lim_{\delta\to0^+}\int_0^1
    \bigg[\mathcal R_\omega(z;r_0)f_\omega(z;r_0)
    \\
    - \frac{\mathcal R_\omega(0;r_H)}{\tilde{\Delta}_H
    \sqrt{(\mathcal Q_H - 1)r_Hz
    \left[2\delta + (1 + \mathcal Q_H)r_Hz\right]}}\bigg]\,\mathrm{d}z\,.
\label{bhatR}
\end{multline}
The result \eqref{polarphase} completes the leading-order description
of quasi-equatorial motion in the strong-deflection regime. Building on
this result and on the equatorial lens equation derived above, we now
turn to the computation of the magnification of higher-order images.

\section{Magnification and caustics}
\label{sec:magncaustics}

Here we compute the magnification of higher-order images and determine the associated caustic points. The magnification is defined as the ratio between the elementary angular area subtended by an image on the observer's sky and the angular area that the unlensed source would subtend at the observer. In the strictly equatorial treatment of the previous sections, the lens map was effectively reduced to one dimension, since both the source and the image were described only by their equatorial angular positions. To compute magnifications, however, one must restore the second angular direction on the observer's sky.

In the quasi-equatorial approximation, the second angular direction is described by small displacements away from the equatorial plane. At leading order in these displacements, the backreaction of the polar motion on the equatorial lens equation can be neglected. The source azimuth $\varphi_S$ is therefore still determined by the equatorial lens equation derived in Sec.~\ref{sec:higherorderimages}, while the source height $h_S$ is determined by the polar motion.

On the image side, a ray is labeled by the angular position $\theta$ in the equatorial direction and by the initial inclination $\Psi_0$, which determines how the photon leaves the observer's sky away from the equatorial plane. Thus, to leading order, the lens map takes the form
\begin{equation}
    (\theta,\Psi_0)\longmapsto (\varphi_S,h_S)\,.
\end{equation}
The magnification is determined by the Jacobian $\mathcal{J}$ of the lens map. Because the equatorial and polar directions decouple at leading order, the Jacobian factorizes into a horizontal derivative, obtained from the equatorial lens equation, and a vertical derivative, obtained from the polar lens equation. The latter reads \cite{Bozza2003}
\begin{multline}
    h_S = h_O\left(\frac{D_{LS}}{\bar{u}}S - C\right) 
    \\
    -\,\Psi_0\left[\left(D_{OL}+D_{LS}\right)C - \frac{D_{OL}D_{LS}}{\bar{u}}S\right]\,,
\label{polarlensequation}
\end{multline}
where $h_O$ is the observer height, kept fixed, and
\begin{equation}
    S \coloneqq \sin(\bar{\varphi}_f)\,,
    \quad
    C \coloneqq \cos(\bar{\varphi}_f)\,.
\end{equation}
For sources and observers far from the black hole, the magnification of the $n$-th image on the prograde branch takes the form \cite{Bozza2003}
\begin{equation}
    \mu = \frac{(D_{OL} + D_{LS})^2}{D_{LS}}\,\frac{1}{|\mathcal{J}|}\,, \quad \mathcal{J} = \frac{\partial \varphi_S}{\partial \theta}\,\frac{\partial h_S}{\partial \Psi_0}\,.
\end{equation}
The relevant derivatives can now be computed from the polar lens equation \eqref{polarlensequation} and from the image-position formula \eqref{sn_improved}. Keeping the derivative obtained from the latter through order $\ell_n^{-3}$, we find
\begin{equation}
    \frac{\partial h_S}{\partial \Psi_0} = -\frac{\mathcal{Y}}{\bar{u}}\,, \quad \frac{\partial \varphi_S}{\partial \theta} = \frac{1}{\theta_\infty \Xi}\,,
\end{equation}
where
\begin{align}
    \mathcal{Y} &\coloneqq \bar{u} \left(D_{OL} + D_{LS}\right)C - D_{OL}D_{LS}S\,,
    \\
    \Xi &\coloneqq \frac{\tilde{a}_1}{\ell_n^2} + \frac{1}{\ell_n^3}\left[\tilde{a}_1 \bar{a}_2 \left(2\log\left(\frac{\ell_n}{\bar{a}_1}\right) - 1\right)\right. 
    \nonumber\\
    &\left.\qquad\qquad\qquad\qquad\qquad\qquad+\, \frac{J_{2,H}}{J_H}\bar{a}_1^2\right]\,.
\end{align}
The formula for the magnification then becomes
\begin{equation}
    \mu = \frac{(D_{OL} + D_{LS})^2}{D_{LS}}\,\frac{\bar{u}\,\theta_\infty |\Xi|}{|\mathcal{Y}|}\,.
\label{magnification}
\end{equation}
We now identify the source positions at which the Jacobian of the lens map vanishes and, correspondingly, the magnification diverges. These locations are the caustic points, i.e., the equatorial-plane restriction of the caustic curves. For the full caustic structure in Kerr lensing, see Ref.~\cite{Bozza2008}. From Eq.~\eqref{magnification}, the caustics are determined by the following condition:
\begin{equation}
    \mathcal{Y} = 0 \quad \Rightarrow \quad S = \frac{\bar{u} \left(D_{OL} + D_{LS}\right)}{D_{OL}D_{LS}}\,C\,.
\end{equation}
Using the definitions of $S$ and $C$, the above equation can be written as
\begin{equation}
    \tan\bar{\varphi}_f =
    \frac{\bar{u}(D_{OL} + D_{LS})}{D_{OL}D_{LS}}\,.
\end{equation}
Since $\bar{u}$ is of the order of the critical impact parameter, while $D_{OL}$ and $D_{LS}$ are much larger in the lensing configuration considered here, the right-hand side is small. Therefore, at leading order in this large-distance expansion, the caustic condition reduces to
\begin{equation}
    \bar{\varphi}_f \simeq k \pi\,, \quad k \in \mathbb{Z}\,.
\label{conditioncaustic}
\end{equation}
Substituting the leading polar phase \eqref{polarphase} and the leading image-position solution
$\delta_n \simeq \bar{a}_1/\ell_n$, Eq.~\eqref{conditioncaustic} gives
\begin{equation}
    \hat{a} \log\left(\frac{\ell_n}{\bar{a}_1}\right) + \hat{b} = k \pi\,.
\end{equation}
Solving this equation for $\ell_n$ gives
\begin{equation}
    \ell_n = \ell_{n,k} = \bar{a}_1 \exp\left(\frac{k \pi - \hat{b}}{\hat{a}}\right)\,.
\end{equation}
Therefore, the azimuthal position of the source at the $k$-th caustic on the $n$-th image branch is
\begin{equation}
    \varphi_S = \varphi_{S,n,k} = 2 \pi n - \bar{b} - \bar{a}_1 \exp\left(\frac{k \pi - \hat{b}}{\hat{a}}\right)\,.
\label{leadingcausticlaw}
\end{equation}
Following Ref.~\cite{Bozza2003}, it is convenient to absorb the winding number $n$ into an unwrapped source azimuth. We therefore introduce $\Phi_S\in(-\infty,\pi]$, whose value modulo $2\pi$ gives the physical source azimuth $\varphi_S\in[-\pi,\pi]$:
\begin{equation}
    \varphi_S = \Phi_S \mod 2\pi\,.
\end{equation}
For the $n$-th higher-order branch, this amounts to choosing
\begin{equation}
    \Phi_S = \varphi_S - 2\pi n\,,
\end{equation}
and therefore Eq.~\eqref{leadingcausticlaw} becomes
\begin{equation}
    \Phi_{S,n,k} = -\bar{b} - \bar{a}_1 \exp\left(\frac{k \pi - \hat{b}}{\hat{a}}\right)\,.
\label{leadingcausticlaw2}
\end{equation}
The above expression represents the leading extremal analog of Eq.~(93) in Ref.~\cite{Bozza2003}. As already noted there, one must restrict to $k\ge2$. Indeed, $k=1$ corresponds to the weak-field caustic, associated with a trajectory whose total angular shift is of order $\pi$, whereas the derivation above relies on the strong-deflection expansion near the critical orbit and is therefore consistent only for relativistic caustics, i.e., those associated with higher-order images produced by photons winding around the black hole.

In summary, whereas in the non-extremal case the caustic positions depend linearly on $k$, in the extremal case they obey an exponential law. This difference is a direct consequence of the near-critical behavior of the two relevant phases. The equatorial extremal lens equation gives $\delta_n\sim \ell_n^{-1}$, while the accumulated polar phase remains logarithmic in $\delta_n$. Solving the caustic condition therefore produces an exponential dependence on $k$.

\subsection{Caustics in extremal Kerr lensing}
\label{sec:causticsKerr}

Let us now specialize Eq.~\eqref{leadingcausticlaw2} to the extremal Kerr spacetime. The strong deflection limit coefficients entering the leading caustic law are
\begin{align}
    \bar{a}_1 &\approx 1.73\,,
    \\
    \bar{b} &\approx -0.95\,,
    \\
    \hat{a} &= 2\,,
    \\
    \hat{b}_D &\approx 2.20\,,
    \\
    \hat{b}_R &\approx 0.95\,,
    \\
    \hat{b} &\approx 3.15\,.
\end{align}
Therefore, for extremal Kerr, Eq.~\eqref{leadingcausticlaw2} becomes
\begin{equation}
    \Phi_{S,n,k} \approx 0.95 - 0.36\exp(1.57 k)\,.
\end{equation}
We can refine this leading law by including the first subleading correction to $\delta_n$. Inserting Eq.~\eqref{deltansubleading} into the caustic condition \eqref{conditioncaustic}, we obtain
\begin{align}
    k \pi &= -\hat{a}\log(\delta_n) + \hat{b}
    \nonumber\\
    &= -\hat{a}\log\left[\frac{\bar{a}_1}{\ell_n}\left(1 + \frac{\bar{a}_2}{\ell_n}\log\left(\frac{\ell_n}{\bar{a}_1}\right)\right)\right] + \hat{b}\,,
\end{align}
which gives
\begin{equation}
    \Phi_{S,n,k} = -\bar{b} - \bar{a}_1 \exp\left(\frac{k \pi - \hat{b}}{\hat{a}}\right) - \frac{\bar{a}_2}{\hat{a}}\left(k\pi - \hat{b}\right)\,.
\label{subleadingcausticlaw}
\end{equation}
Recalling that, for extremal Kerr, $\bar{a}_2=4\sqrt{3}/9$, Eq.~\eqref{subleadingcausticlaw} yields
\begin{equation}
    \Phi_{S,n,k} \approx 2.17 - 1.21k - 0.36\exp(1.57 k)\,.
\label{subleadingcausticlaw2}
\end{equation}
A further improvement is obtained by including the first subleading correction to the polar phase. The derivation is not reported explicitly, since it follows the same endpoint-expansion strategy used throughout the paper:~one expands the smooth factors of the polar-phase integrand one order beyond the leading approximation, while postponing the small-$\delta$ expansion of the endpoint-sensitive terms until after integration. This leads to
\begin{equation}
    \bar\varphi_f(\delta) \simeq
    -\hat{a}\log(\delta) + \hat{b} + \delta\left[-\hat a_1\log(\delta) + \hat b_1\right]\,,
\end{equation}
where the quantities $\hat a$ and $\hat b$ are the leading-order coefficients of Eq.~\eqref{polarphase}. The coefficients $\hat{a}_1$ and $\hat{b}_1$ are determined by the next-to-leading endpoint expansion of Eqs.~\eqref{Romega} and \eqref{fomega}, including the $\delta$-dependence of $\bar u$ through $u\simeq J(\delta)$; the coefficient $\hat{b}_1$ also receives a contribution from the regular part of the integral. For extremal Kerr, we find
\begin{align}
    &\hat{a}_1 = 0\,,
    \\
    &\hat{b}_1 = \frac{16}{9}\,.
\end{align}
As a result, the caustic law \eqref{subleadingcausticlaw2} receives the additional constant contribution
\begin{equation}
    \frac{\bar{a}_1\hat{b}_1}{\hat{a}} =
    \frac{8\sqrt{3}}{9}\,,
\end{equation}
and becomes
\begin{equation}
    \Phi_{S,n,k} \approx 3.71 - 1.21 k - 0.36\exp(1.57k)\,.
\label{subleadingcausticlaw4}
\end{equation}

We now compare our results with the ones of Ref.~\cite{Bozza2008}. We first account for the different choice of azimuthal origin. In the present work the angular coordinate is fixed so that $\varphi_O=\pi$, whereas the convention used in Ref.~\cite{Bozza2008} sets $\varphi_O=0$. Therefore, the corresponding source azimuth is obtained by subtracting $\pi$ from Eq.~\eqref{subleadingcausticlaw4}:
\begin{equation}
    \Phi_{S,n,k} - \pi \approx 0.56 - 1.21k - 0.36\exp(1.57k)\,.
\end{equation}
The exponential contribution agrees with the corresponding term in Eq.~(16) of Ref.~\cite{Bozza2008}. The remaining difference lies in the non-exponential part:~our expression contains a constant plus a linear term, $0.56-1.21k$, whereas the fit presented in Eq.~(16) of Ref.~\cite{Bozza2008} is purely linear in $k$. Over the finite range $k=2,3,4,5$, the quantity $0.56-1.21k$ can be approximated by a purely linear term $-b_{\rm eff}k$. A least-squares fit gives
\begin{align}
    &b_{\rm eff} =
    -\frac{\sum_{k=2}^{5} k\,y_k}{\sum_{k=2}^{5} k^2}
    \approx 1.06\,,
    \\
    &y_k \coloneqq 0.56 - 1.21k\,.
\end{align}
This is in very good agreement with the coefficient $1.05$ appearing in Eq.~(16) of Ref.~\cite{Bozza2008}.

\section{Summary and conclusions}
\label{sec:conclusions}

In this work we have developed a strong deflection limit expansion for a class of extremal rotating black holes in which the equatorial prograde critical photon orbit coincides with the degenerate event horizon. This situation is qualitatively different from the standard non-extremal case. Because the critical orbit lies at the horizon, the turning-point singularity of the deflection integral overlaps with the near-horizon singular structure of the metric. As a consequence, the deflection angle is no longer dominated only by a logarithmic divergence: its leading behavior contains a power-law term proportional to $1/\delta$, accompanied by a subleading logarithmic contribution.

We first formulated the expansion for a generic stationary, axisymmetric, asymptotically flat metric satisfying the extremal near-horizon conditions discussed in Sec.~\ref{sec:prelanddeflangle}. We derived the leading strong deflection coefficients and then extended the calculation by retaining terms linear in the near-critical parameter $\delta=r_0-r_H$. This improved expansion is important for describing the first higher-order images, which in the extremal case are pushed farther from the shadow edge with respect to the non-extremal case.

We then applied the general formalism to explicit extremal rotating geometries. For extremal Kerr, all coefficients entering the improved deflection angle can be obtained analytically. For extremal Kerr-Newman, we identified the region of the extremal parameter space in which the prograde critical radius coincides with the horizon, namely $a\ge 1/4$, or equivalently $|Q|\le \sqrt{3}/4$. Within this regime, we computed the strong deflection coefficients and studied the dependence of the image positions on the charge. We also considered extremal Kerr-Sen black holes, for which the prograde critical radius coincides with the horizon throughout the extremal family. These examples show how the additional parameters of the spacetime affect both the strong deflection coefficients and the location of the higher-order images.

Finally, we extended the analysis beyond strictly equatorial motion by considering quasi-equatorial photon trajectories. In this regime, the polar motion can be treated perturbatively while the equatorial lens equation remains unchanged at leading order. We derived the accumulated polar phase in the strong deflection limit and used it to construct the quasi-equatorial lens map. This allowed us to compute the magnification of higher-order images and to determine the associated caustic points. In particular, we found that the caustic positions obey an exponential law in the caustic index $k$. This behavior is the extremal analog of the linear caustic structure found in the non-extremal case. Moreover, for extremal Kerr, the resulting caustic law agrees with the known numerical results obtained in Ref.~\cite{Bozza2008}.

Some extensions of this work are quite natural. First, throughout the present analysis the source and the observer have been treated in the asymptotic region. For astrophysical applications, however, the source may be located close to the black hole, and finite-distance effects are important. It would therefore be interesting to generalize the present extremal strong deflection expansion to sources at finite distance, following the type of formalism developed in Ref.~\cite{BozzaScarpetta2007}.

A second important direction is the extension to near-extremal black holes. Although exactly extremal Kerr black holes are special, this does not necessarily prevent a smooth approach to extremality at the level of lensing observables. Evidence from near-extremal Kerr indicates that, after an appropriate scaling of the near-horizon variables, the corresponding lensing problem admits a smooth limit to the extremal case \cite{Gralla2018}. This provides a natural motivation for extending the present formalism to rapidly rotating but subextremal black holes, where one expects a crossover description interpolating between the standard logarithmic behavior and the extremal asymptotics derived here.

Finally, it would certainly be interesting to go beyond the quasi-equatorial approximation and develop a fully two-dimensional lensing treatment. This would require allowing generic photon trajectories, including the backreaction of the polar motion on the equatorial dynamics, and would provide a more complete description of the image structure and caustics around extremal/near-extremal rotating black holes.

\section*{Acknowledgements} \label{sec:acknowledgements}

FF gratefully acknowledges the hospitality of the Center of Gravity at the Niels Bohr Institute.

\appendix

\onecolumngrid

\section{Leading endpoint expansion of the deflection integral}
\label{app:powerlawdivergence}

In this appendix, we derive the leading singular approximation used in
Eq.~\eqref{azshiftintegral}. The divergence is controlled by the endpoint $z = 0$ in the near-critical limit \(\delta\to0^+\), with $\delta$ defined in Eq.~\eqref{delta}. We therefore retain the $z$- and $\delta$-dependence
of the factors that control the boundary region $z \sim \delta$. By contrast, all prefactors that remain regular as $z \to 0$ and $r_0 \to r_H$ can be evaluated directly at the horizon.

We start from Eq.~\eqref{integrand2}, namely
\begin{equation}
    \frac{\mathrm{d}\varphi}{\mathrm{d}z} = \frac{r_0\,\tilde{\mathcal B}(r(z))
    \left[\mathcal D(r(z)) - J\right]
    }{\left(\delta + r_H z\right)^2 \tilde{\Delta}(r(z))\sqrt{\mathcal V(r(z);r_0)}}\,,
\label{spappendix}
\end{equation}
where we recall that
\begin{equation}
    r(z) = \frac{r_0}{1 - z}\,,
    \quad
    \delta = r_0 - r_H\,,
    \quad
    J = \mathcal D_0 - \tilde{\Delta}_0\,\delta\,,
    \quad
    \mathcal V(r(z);r_0) =
    \mathcal C(r(z))^2 + J\left[J - 2\mathcal D(r(z))\right]\,.
\end{equation}
Let us first consider the factors that are innocuous in the endpoint
region. At leading order, we write
\begin{equation}
    r_0 \frac{\tilde{\mathcal B}(r(z))}{\tilde{\Delta}(r(z))} \simeq r_H
    \frac{\tilde{\mathcal B}_H}{\tilde{\Delta}_H}\,.
\end{equation}
We now turn to the quantity $\mathcal D(r(z))-J$ in Eq.~\eqref{spappendix}. Using the expression for $J_- = J$ in Eq.~\eqref{J}, we have
\begin{equation}
    \mathcal D(r(z)) - J =
    \mathcal D(r(z)) - \mathcal D_0 + \tilde{\Delta}_0\,\delta\,.
\end{equation}
Expanding first in $z$ gives
\begin{equation}
    \mathcal D(r(z)) - \mathcal D_0 = \mathcal D'_0\left(r(z) - r_0\right) + \mathcal{O}(z^2) = \mathcal D'_0r_0z + \mathcal{O}(z^2)\,.
\end{equation}
The coefficients multiplying the endpoint-sensitive combination can now
be safely evaluated at $r_0=r_H$, yielding
\begin{equation}
    \mathcal D(r(z)) - J \simeq
    \tilde{\Delta}_H\,\delta + \mathcal D'_Hr_Hz\,.
\end{equation}
For later convenience, using the quantity $\mathcal Q_H$ defined in the main text, we rewrite the above equation as
\begin{equation}
    \mathcal D(r(z)) - J \simeq
    \tilde{\Delta}_H\left(\delta + \mathcal Q_Hr_Hz\right)\,, \quad \mathcal Q_H = \frac{\mathcal D'_H}{\tilde{\Delta}_H} = \frac{\mathcal C'_H}{\tilde{\Delta}_H}\,.
\label{numleadingAppA}
\end{equation}
We proceed by expanding the square-root argument in Eq.~\eqref{spappendix}. We first rewrite $\mathcal V(r(z);r_0)$ as
\begin{equation}
    \mathcal V(r(z);r_0) = \left(
    \mathcal D(r(z)) - J\right)^2
    - \left(\mathcal D(r(z))^2 - \mathcal C(r(z))^2\right)\,.
\end{equation}
The expansion of the first term is obtained from Eq.~\eqref{numleadingAppA}. As for the
second term, using Eq.~\eqref{Deltaconformal}, the leading endpoint-sensitive contribution is given by
\begin{equation}
    \mathcal D(r(z))^2-\mathcal C(r(z))^2
    \simeq
    \tilde{\Delta}_H^2
    \left(\delta + r_Hz\right)^2.
\end{equation}
Therefore, $\mathcal{V}(r(z);r_0)$ is finally written as
\begin{align}
    \mathcal V(r(z);r_0) &\simeq
    \tilde{\Delta}_H^2\left[\left(\delta + \mathcal Q_Hr_Hz\right)^2 -
    \left(\delta + r_H z\right)^2\right]
    \nonumber\\
    &= \tilde{\Delta}_H^2
    (\mathcal Q_H - 1)r_Hz\left[
    2\delta + (1 + \mathcal Q_H)r_H z\right]\,.
\end{align}
Putting the pieces together, the leading endpoint approximation of the
integrand is
\begin{equation}
    \frac{\mathrm{d}\varphi}{\mathrm{d}z} \simeq r_H\frac{\tilde{\mathcal B}_H}{\tilde{\Delta}_H}\frac{\delta + \mathcal Q_Hr_Hz}{(\delta + r_H z)^2\sqrt{(\mathcal Q_H - 1)r_Hz
    \left[2\delta + (1 + \mathcal Q_H)r_Hz\right]}}\,.
\end{equation}
Multiplying by the factor $2$ that accounts for the ingoing and
outgoing parts of the trajectory gives precisely the singular
contribution displayed in Eq.~\eqref{azshiftintegral}.

\section{Subleading endpoint expansion of the deflection integral}
\label{app:logdivergence}

In this appendix, we derive the first subleading correction to the
endpoint approximation obtained in Appendix~\ref{app:powerlawdivergence}.
This correction is responsible for the logarithmic divergence in the
deflection angle. The strategy is the same as in Appendix~\ref{app:powerlawdivergence}. We start again from Eq.~\eqref{spappendix}. We first consider the regular prefactor. Expanding it around $z=0$ gives
\begin{equation}
    \frac{\tilde{\mathcal B}(r(z))}{\tilde{\Delta}(r(z))} \simeq \frac{\tilde{\mathcal B}_H}{\tilde{\Delta}_H}\left[1 
    + r_H\left(\frac{\tilde{\mathcal B}'_H}{\tilde{\mathcal B}_H} -\frac{\tilde{\Delta}'_H}{\tilde{\Delta}_H}
    \right)z\right]\,.
\label{prefactorAppB}
\end{equation}
Here, as throughout this appendix, the coefficients of the subleading
terms have been evaluated at $\delta=0$, since they are smooth in the endpoint region. Next, the endpoint expansion of $\mathcal D(r(z))-J$ results in
\begin{equation}
    \mathcal D(r(z)) - J \simeq \tilde{\Delta}_H\left(\delta + \mathcal Q_H r_H z\right) + \tilde{\Delta}_H\mathcal Q_H r_H\left(
    1 + \frac{r_H\mathcal D''_H}{2\mathcal D'_H}\right)z^2\,.
\label{numeratorAppB}
\end{equation}
Finally, the expansion of the inverse square-root factor yields
\begin{equation}
    \frac{1}{\sqrt{\mathcal V(r(z);r_0)}} \simeq \frac{1}{\tilde{\Delta}_H\sqrt{(\mathcal Q_H - 1)r_Hz\left[2\delta + (1 + \mathcal Q_H)r_Hz\right]}}\left[1 - \left(1 + \frac{r_H\mathcal V_3}{2\mathcal V_2}\right)z\right]\,,
\label{sqrtInverseAppB}
\end{equation}
with $\mathcal V_2$ and $\mathcal V_3$, already defined in the main text, given by
\begin{equation}
    \mathcal V_2 = \mathcal D_H^{\prime 2} - \tilde{\Delta}_H^2\,,
    \quad
    \mathcal V_3 = \mathcal D'_H\mathcal D''_H - 2\tilde{\Delta}_H\tilde{\Delta}'_H\,.
\label{V2V3appendix}
\end{equation}
We can now multiply Eqs.~\eqref{prefactorAppB},
\eqref{numeratorAppB}, and \eqref{sqrtInverseAppB}. Keeping only the terms that multiply the leading endpoint kernel at relative order $z$,
we obtain
\begin{equation}
    \frac{\mathrm{d}\varphi}{\mathrm{d}z} \simeq r_H \frac{\tilde{\mathcal B}_H}{\tilde{\Delta}_H}\frac{\delta + \mathcal Q_H r_H z}{(\delta + r_H z)^2\sqrt{(\mathcal Q_H - 1)r_H z
    \left[2\delta + (1 + \mathcal Q_H)r_H z\right]}}\left(1 + qz\right)\,,
\end{equation}
where the correction coefficient is read off as
\begin{equation}
    q = r_H\left(\frac{\tilde{\mathcal B}'_H}{\tilde{\mathcal B}_H} - \frac{\tilde{\Delta}'_H}{\tilde{\Delta}_H}\right) + \frac{r_H}{2}\left(\frac{\mathcal D''_H}{\mathcal D'_H} - \frac{\mathcal V_3}{\mathcal V_2}\right)\,.
\end{equation}
This is the coefficient used in Eq.~\eqref{defq} of the main text.

\section{Calculation of the regular remainder in Kerr via matched asymptotic expansion}
\label{app:maeKerr}

As already anticipated, in the extremal case the evaluation of the deflection integral involves a non-uniform small-$\delta$ limit, so that a naive Taylor expansion in $\delta$ at fixed $z$ cannot be interchanged with the $z$-integration. Here we will focus on the calculation of $\bar{b}_R$. Schematically, let us consider the integral
\begin{equation}
    I(\delta) = \int_0^1 f(z,\delta)\,\mathrm{d}z\,,
\label{eq:Idef-general}
\end{equation}
where the integrand contains $\delta$ through combinations such as $(z\,+\,\lambda_1\,\delta)^{-2}$ and $\sqrt{z(z + \lambda_2\,\delta)}$, with $\lambda_1$, $\lambda_2$ finite constants. Expanding at fixed $z$, we would implicitly assume $\delta/z\ll 1$; for instance,
\begin{equation}
    \frac{1}{(z + \lambda_1\,\delta)^{2}} = \frac{1}{z^{2}}\left(1 + \frac{\lambda_1\,\delta}{z}\right)^{-2}
    = \frac{1}{z^{2}}\left[1 - 2\frac{\lambda_1\,\delta}{z} + 3\left(\frac{\lambda_1\,\delta}{z}\right)^{2} + \cdots\right]\,,
\label{seriesexample}
\end{equation}
which is valid only for $z \gg  \delta$. Since the integration domain in Eq.~\eqref{eq:Idef-general} includes the boundary layer $z\sim \delta$, the series~\eqref{seriesexample} is not uniformly valid on $z\in[0,1]$, and term-by-term integration would lead to an incorrect result. One way to proceed is to use a matched asymptotic expansion; we first introduce
\begin{equation}
    z_{*} = \delta^{m}, \qquad 0 < m < 1,
\label{eq:zstar}
\end{equation}
and then split the integral \eqref{eq:Idef-general} as
\begin{align}
    I(\delta) &= I_{\mathrm{in}}(\delta) + I_{\mathrm{out}}(\delta)\,, \\
    I_{\mathrm{in}}(\delta) &\coloneqq \int_0^{z_{*}} f(z,\delta)\,\mathrm{d}z\,, \\
    I_{\mathrm{out}}(\delta) &\coloneqq \int_{z_{*}}^1 f(z,\delta)\,\mathrm{d}z\,.
\end{align}
As far as $I_{\text{in}}(\delta)$ is concerned, we proceed by changing variables to
\begin{equation}
    z = \delta\,y\,,
\end{equation}
which immediately yields
\begin{equation}
    I_{\mathrm{in}}(\delta) = \int_0^{z_{*}/\delta} \delta\,f(\delta\,y,\delta)\,\mathrm{d}y = \int_0^{\delta^{m-1}} \delta\,f(\delta\,y,\delta)\,\mathrm{d}y\,.
\label{eq:Iin}
\end{equation}
The choice \(z_* = \delta^{m}\) with \(0 < m < 1\) places the splitting point in the intermediate range \(\delta \ll z_* \ll 1\) as \(\delta \to 0^+\). Indeed, since \(z_* = \delta^m \to 0\) as \(\delta \to 0^+\) (\(m>0\)), the $\delta \to 0^+$ asymptotics of the outer integral is sensitive to the behavior near its lower endpoint within \(\delta \ll z \ll 1\); as for the inner integral, we have that
\begin{equation}
    y_* = \frac{z_*}{\delta} = \delta^{m-1} \to \infty \,\,\,\, \text{as} \,\,\,\, \delta \to 0^+ \,\,\,\, (m < 1)\,,
\end{equation}
so \(z_*\) is much larger than \(\delta\). In other words, the inner description covers not only \(y = \mathcal{O}(1)\) (i.e., \(z \sim \delta\)), but also \(y \gg 1\), which corresponds to \(z \gg \delta\). Consequently, the inner approximation (large $y=z/\delta$) and the outer approximation (small $\delta/z$) are simultaneously valid in the overlap regime $\delta \ll z \ll 1$ and, when truncated at the same order, agree on a common asymptotic expansion of the exact integrand. Therefore, any explicit $z_*$-dependence produced by the split can only enter through terms evaluated at the splitting point when the overlap expansion is integrated. Such $z_*$-dependent terms (e.g., $\log z_*$ or powers of $z_*$) appear with opposite signs in $I_{\mathrm{in}}(\delta)$ and $I_{\mathrm{out}}(\delta)$ and cancel in $I(\delta) = I_{\mathrm{in}}(\delta) + I_{\mathrm{out}}(\delta)$, leaving an expansion independent of $m$ up to the truncation error.

We can now safely expand the inner integrand because the rescaling \(z = \delta y\) absorbs the non-uniform dependence on the ratio \(\delta/z\):~on the inner domain \(z = \mathcal{O}(\delta)\) the shifted factors become \((z + \lambda\,\delta) = \delta(y + \lambda)\) (with $\lambda$ being a finite constant) and \(\delta/z = 1/y\), so that the scaled integrand \(F(y,\delta) \equiv \delta f(\delta y,\delta)\) admits an expansion for small \(\delta\) for each fixed \(y\), without requiring \(\delta/z \ll 1\). We then integrate the resulting coefficient functions in \(y\) up to \(y_{*} = \delta^{m - 1}\), thereby retaining the full overlap/tail contribution (e.g., terms \(\propto \log y_{*} \sim  \log\delta\)). Naturally, term-by-term integration of the inner \(\delta\)-expansion is then justified provided the integrated remainder satisfies
\begin{equation}
    \int_0^{y_*} R_N(y,\delta)\,\mathrm{d}y = o\bigl(\delta^N\bigr) \,\,\,\, \text{as} \,\,\,\, \delta \to 0^+\,,
\end{equation}
where \(N\ge0\) is the highest order kept in the integrand expansion (this condition can be easily verified a posteriori for the specific integrand). In other words, integrating to $y_* = \delta^{m - 1} \to \infty$ as $\delta \to 0^+$ is fine provided the growth of the integration range does not magnify the remainder. In the outer region \(z\in[z_{*},1]\) we instead have a uniform small parameter, namely
\begin{equation}
    \frac{\delta}{z}\le\frac{\delta}{z_{*}} = \delta^{1 - m} \to 0,
\end{equation}
so we can safely expand the integrand at fixed \(z\) and then integrate term by term on \([z_{*},1]\).

Let us now apply this strategy to compute the strong deflection limit coefficient $\bar{b}_R$ in the Kerr case. From Eq.~\eqref{regularpart}, we obtain
\begin{equation}
    \varphi_{f,R}(\delta) = \int_0^1 \left(4\sqrt{\frac{(z + \delta)^2(1 + 2\delta)^3}{z(z + 2\delta)^4 \left[z(3 - 2z) + 2\delta(2 - z)\right]}} - \frac{4(3 + z)(z + \delta)}{3(z + 2\,\delta)^2 \sqrt{z(3z + 4\,\delta)}}\right) \mathrm{d}z \coloneqq \int_0^1 g(z,\delta)\,\mathrm{d}z\,.
\end{equation}
Now, we split the above integral as
\begin{align}
    \varphi_{f,R}(\delta) &= \varphi_{f,R}^\mathrm{in}(\delta) + \varphi_{f,R}^\mathrm{out}(\delta)\,, \\
    \varphi_{f,R}^\mathrm{in}(\delta) &\coloneqq \int_0^{z_{*}} g(z,\delta)\,\mathrm{d}z\,, \\
    \varphi_{f,R}^\mathrm{out}(\delta) &\coloneqq \int_{z_{*}}^1 g(z,\delta)\,\mathrm{d}z\,,
\end{align}
where $z_*$ is chosen as in Eq.~\eqref{eq:zstar}, with $m = 1/2$. We proceed by computing $\varphi_{f,R}^\mathrm{in}(\delta)$. Changing variables to $z = \delta\,y$, we can write
\begin{equation}
    \varphi_{f,R}^\mathrm{in}(\delta) = \int_0^{z_{*}/\delta} \delta\,g(\delta\,y,\delta)\,\mathrm{d}y = \int_0^{1/\sqrt{\delta}} \delta\,g(\delta\,y,\delta)\,\mathrm{d}y\,.
\end{equation}
Expanding $\delta\,g(\delta\,y,\delta)$ about $\delta = 0$ and retaining the leading term yields
\begin{equation}
    \delta\,g(\delta\,y,\delta) \simeq \frac{8(1 + y)\left(18 + 13y\right)}{3\sqrt{y}(y + 2)^2(3y + 4)^{3/2}}\,.
\end{equation}
Integrating the above expression from $y = 0$ to $y = 1/\sqrt{\delta}$ and expanding the result about $\delta = 0$ gives
\begin{equation}
    \varphi_{f,R}^\mathrm{in}(\delta) \simeq \frac{2}{9}\left(11\sqrt{3} + 6 \operatorname{arcoth}\!\big(\sqrt{3} \big)\right) \equiv \bar{b}_{R,\text{in}}\,.
\label{bRin}
\end{equation}
Let us now compute $\varphi_{f,R}^\mathrm{out}(\delta)$. Expanding $g(z,\delta)$ and keeping only the leading-order term gives
\begin{equation}
    g(z,\delta) \simeq 4\,\frac{9(3 - 2z)^{-1/2} - 3\sqrt{3} - \sqrt{3}z}{9z^2}\,.
\label{gouter}
\end{equation}
Integrating Eq.~\eqref{gouter} from $z = \sqrt{\delta}$ to $z = 1$ and then expanding about $\delta = 0$ results in
\begin{equation}
    \varphi_{f,R}^\mathrm{out}(\delta) \simeq \frac{2}{9}\left[4\sqrt{3}\left(1 + \log\big(3 - \sqrt{3}\big)\right) - 6\right] \equiv \bar{b}_{R,\text{out}}\,.
\label{bRout}
\end{equation}
Adding Eqs.~\eqref{bRin} and \eqref{bRout} finally leads to
\begin{equation}
    \bar{b}_R = \bar{b}_{R,\text{in}} + \bar{b}_{R,\text{out}} = \frac{2}{9}\left(4\sqrt{3}\log\left(3 - \sqrt{3}\right) + 6\operatorname{arcoth}\!\left(\sqrt{3} \right) - 6 + 15\sqrt{3}\right)\,.
\end{equation}
This is the result presented in the main text. Note that the cutoff $z_*$ can enter in two qualitatively different ways. 
First, the overlap contribution may generate endpoint-singular terms (power-law, logarithmic), which are removed by construction in the definition of the finite remainder coefficient $\bar{b}_R$. Second, there are genuinely artificial $z_*$-dependent contributions that are integrable at $z = 0$ and therefore appear as positive powers of $z_*$; after setting $z_*=\delta^m$, these become higher-order terms in $\delta$ and may be discarded if one truncates the small-$\delta$ expansions of $I_{\mathrm{in}}(\delta)$ and $I_{\mathrm{out}}(\delta)$ separately, so their cancellation is not necessarily visible term-by-term. Let us try to be more explicit about this. We first consider the inner piece. Define $F_{\text{in}}(y,\delta) \equiv \delta\,g(\delta\,y,\delta)$. Expanding to leading order about $\delta = 0$, we have
\begin{equation}
    F_{\text{in}}(y,\delta) \simeq F_0(y)\,.
\end{equation}
Then, we can write
\begin{equation}
    \int_0^{\delta^{m - 1}} F_0(y)\,\mathrm{d}y = \int_0^\infty F_0(y)\,\mathrm{d}y - \int_{\delta^{m - 1}}^\infty F_0(y)\,\mathrm{d}y = \frac{2}{9}\left(11\sqrt{3} + 6 \operatorname{arcoth}\!\left(\sqrt{3} \right)\right) - \int_{\delta^{m - 1}}^\infty F_0(y)\,\mathrm{d}y\,.
\end{equation}
The only $m$-dependence lies in the tail from $y_* = \delta^{m - 1}$ to $\infty$. Estimating that tail requires the $y \to \infty$ asymptotics of $F_0(y)$ (note that we are not claiming the whole integral is controlled by large $y$, only that the difference on the left-hand side of the first equality in the above expression is controlled by large $y$). Therefore, let us expand $F_0(y)$ for large $y$, obtaining
\begin{equation}
    F_0(y) \simeq \frac{104}{9\sqrt{3}}\frac{1}{y^2} - \frac{376}{9\sqrt{3}}\frac{1}{y^3} + \mathcal{O}\left(y^{-4}\right)\,.
\end{equation}
Therefore, the tail evaluates to
\begin{equation}
    \int_{\delta^{m - 1}}^\infty F_0(y)\,\mathrm{d}y \simeq \frac{104}{9\sqrt{3}}\delta^{1 - m} - \frac{376}{18\sqrt{3}}\delta^{2(1 - m)}\,.
\end{equation}
As we can see, the first $m$-dependent term is proportional to $\delta^{1 - m}$, i.e., beyond leading order. As far as the outer piece is concerned, we write
\begin{equation}
    \int_{\delta^m}^1 g_0(z)\,\mathrm{d}z = \int_0^1 g_0(z)\,\mathrm{d}z - \int_0^{\delta^m} g_0(z)\,\mathrm{d}z = \frac{2}{9}\left[4\,\sqrt{3}\left(1 + \log\left(3 - \sqrt{3}\right)\right) - 6\right] - \int_0^{\delta^m} g_0(z)\,\mathrm{d}z\,,
\end{equation}
where $g_0(z)$ is the leading-order coefficient function in the $\delta$-expansion of $g(z,\delta)$. So the only $m$-dependence lies in the small strip from $z = 0$ to $z = \delta^m$. Expanding $g_0(z)$ about $z = 0$ and evaluating the strip gives
\begin{equation}
    \int_0^{\delta^m} g_0(z)\,\mathrm{d}z \simeq \frac{2}{3\sqrt{3}}\,\delta^m + \frac{5}{27\sqrt{3}}\,\delta^{2m}\,.
\end{equation}
The first $m$-dependent term is proportional to $\delta^m$.

\section{Endpoint expansion at linear order in \texorpdfstring{\(\delta\)}{delta}}
\label{app:linear-delta}

In this appendix, we derive the endpoint expansion needed to obtain the improved deflection angle of Sec.~\ref{sec:improvdeflangle}. At logarithmic order, the leading endpoint integrand was multiplied by the correction $1 + qz$. To reach linear order in $\delta$, we must extend the endpoint expansion by one further order and write the multiplicative correction as
\begin{equation}
    1 + qz + wz^2\,.
\end{equation}
The quantities $q$ and $w$, together with their separate
contributions $q_{\mathcal B\Delta},q_{\mathcal D},q_{\mathcal V}$ and
$w_{\mathcal B\Delta},w_{\mathcal D},w_{\mathcal V}$, were defined in
Sec.~\ref{sec:improvdeflangle}. Here we detail the origin of these coefficients. We proceed as in Appendices~\ref{app:powerlawdivergence} and
\ref{app:logdivergence}:~the endpoint-sensitive leading structure is kept with its full $z$- and $\delta$-dependence, while the
coefficients of the subleading $z$-corrections are evaluated at $\delta=0$. The expansion of the regular prefactor gives
\begin{equation}
    \frac{\tilde{\mathcal B}(r(z))}{\tilde{\Delta}(r(z))} \simeq
    \frac{\tilde{\mathcal B}_H}{\tilde{\Delta}_H}\left(1 + q_{\mathcal B\Delta}z +
    w_{\mathcal B\Delta}z^2\right)\,,
\end{equation}
where
\begin{equation}
    q_{\mathcal B\Delta} =
    r_H\left(\frac{\tilde{\mathcal B}'_H}{\tilde{\mathcal B}_H} -\frac{\tilde{\Delta}'_H}{\tilde{\Delta}_H}\right)\,, \quad w_{\mathcal B\Delta} =
    q_{\mathcal B\Delta} + \frac{r_H^2}{2}\left[\Bigg(\frac{\tilde{\mathcal B}}{\tilde{\Delta}}\Bigg)''\frac{\tilde{\Delta}}{\tilde{\mathcal B}}\right]_H\,.
\end{equation}
Next, the endpoint expansion of the factor $\mathcal D(r(z))-J$ gives
\begin{equation}
    \mathcal D(r(z)) - J \simeq
    \tilde{\Delta}_H\left(\delta + \mathcal Q_H r_H z\right)\left(1 + q_{\mathcal D}z + w_{\mathcal D}z^2\right)\,,
\end{equation}
with
\begin{equation}
    q_{\mathcal D} = 1 + \frac{r_H\mathcal D''_H}{2\mathcal D'_H}\,, \quad  w_{\mathcal D} = 1 + \frac{r_H\mathcal D''_H}{\mathcal D'_H} + \frac{r_H^2\mathcal D'''_H}{6\mathcal D'_H}\,.
\end{equation}
Finally, the expansion of the inverse square-root factor yields
\begin{equation}
    \frac{1}{\sqrt{\mathcal V(r(z);r_0)}} \simeq \frac{1}{\tilde{\Delta}_H\sqrt{(\mathcal Q_H - 1)r_Hz\left[2\delta + (1 + \mathcal Q_H)r_Hz\right]}}
    \left(1 + q_{\mathcal V}z + w_{\mathcal V}z^2\right)\,,
\end{equation}
where
\begin{equation}
    q_{\mathcal V} = -\left(1 + \frac{r_H\mathcal V_3}{2\mathcal V_2}\right)\,, \quad w_{\mathcal V} = r_H^2
    \left(\frac{3\mathcal V_3^2}{8\mathcal V_2^2} - \frac{\mathcal V_4}{2\mathcal V_2}\right)\,.
\end{equation}
In the expression above, $\mathcal V_2$ and $\mathcal V_3$ are given in Eq.~\eqref{V2V3appendix}, while $\mathcal V_4$, already defined in the main text, is
\begin{equation}
    \mathcal V_4 = \frac{\mathcal D_H^{\prime\prime 2}}{4} + \frac{\mathcal D'_H\mathcal D'''_H}{3} - \left(\tilde{\Delta}_H^{\prime 2} + \tilde{\Delta}_H\tilde{\Delta}''_H\right)\,.
\end{equation}
Multiplying the three corrections and expanding up to second order in $z$, we find
\begin{multline}
    \left(1 + q_{\mathcal B\Delta}z + w_{\mathcal B\Delta}z^2\right)\left(1 + q_{\mathcal D}z + w_{\mathcal D}z^2\right)\left(1 + q_{\mathcal V}z + w_{\mathcal V}z^2\right)
    \\
    = 1 + \underbrace{
    \left(q_{\mathcal B\Delta} +
    q_{\mathcal D} + q_{\mathcal V}
    \right)}_{\displaystyle = q}z +
    \underbrace{\left(w_{\mathcal B\Delta} + w_{\mathcal D} + w_{\mathcal V} + q_{\mathcal B\Delta}\,q_{\mathcal D} + q_{\mathcal B\Delta}\,q_{\mathcal V} + q_{\mathcal D}\,q_{\mathcal V}\right)}_{\displaystyle = w}z^2 + \mathcal{O}(z^3)\,.
\end{multline}
This reproduces the coefficient $q$ used in Appendix~\ref{app:logdivergence} and the quadratic coefficient
$w$ entering the improved expansion. Putting everything together, we then obtain
\begin{equation}
    \frac{\mathrm{d}\varphi}{\mathrm{d}z} \simeq r_H \frac{\tilde{\mathcal B}_H}{\tilde{\Delta}_H}\frac{\delta + \mathcal Q_H r_H z}{(\delta + r_H z)^2\sqrt{(\mathcal Q_H - 1)r_H z
    \left[2\delta + (1 + \mathcal Q_H)r_H z\right]}}\left(1 + qz + wz^2\right)\,.
\end{equation}

\twocolumngrid

\end{document}